\documentclass[12pt,english,aps,manuscript]{article}
\usepackage[T1]{fontenc}
\usepackage[latin9]{inputenc}
\usepackage{geometry}
\usepackage{amsmath}
\usepackage{amssymb}
\usepackage{graphicx}
\usepackage{esint}
\usepackage[numbers]{natbib}

\makeatletter
\usepackage{babel}

\makeatother

\usepackage{babel}
\begin{document}
\begin{center}
\textbf{\Large{}{}Revisiting Vacuum decay in Field Theory }{\Large\par}
\par\end{center}

\begin{center}
 
\par\end{center}

\begin{center}
\vspace{0.3cm}
\par\end{center}

\begin{center}
{\large{}S. P. de Alwis$^{\dagger}$ }{\large\par}
\par\end{center}

\begin{center}
Physics Department, University of Colorado, \\
 Boulder, CO 80309 USA 
\par\end{center}

\begin{center}
\textbf{Abstract} 
\par\end{center}

\begin{center}
\vspace{0.3cm}
 
\par\end{center}

\smallskip{}
 \vspace{0.3cm}
 We revisit the formalism for tunneling in quantum field theory developed
by Coleman and collaborators. In particular using the generalization
of WKB methods for tunneling in quantum mechanics we avoid the problems
with negative eigenvalues and convexity issues associated with Coleman's
approach. While the exponential factor is the same, we find differences
in the pre-factor. Then we point out that to actually discuss the
time evolution of the state, we need a wave packet formulation which
we proceed to discuss. Next we address the problem of justifying the
application of semi-classical tunneling calculations to the decay
of the standard model vacuum, where the classical potential signifies
absolute stability, though the effective potential appears to imply
the possibility of meta-stability (with more than one local minimum).
This is in contrast to the usual situation in applications of the
formalism for tunneling, where the \textit{classical} potential has
more than one local minimum. 
\begin{center}
\vspace{0.3cm}
\par\end{center}

\vfill{}

$^{\dagger}$ dealwiss@colorado.edu

\eject

\section{Introduction}

The discussion of vacuum transitions in field theory was initiated
by Coleman\citep{Coleman:1977py,Callan:1977pt,Coleman:1980aw} in
the nineteen seventies\footnote{Actually there are somewhat earlier papers by P. Frampton which addressed
vaccum tunneling \citep{Frampton:1976kf,Frampton:1976pb}, however
it seems that most of the subsequent literature was based on Coleman's
discussion.}. It is based on the following argument. Consider a field theory for
a scalar field $\phi$, with a potential $V(\phi)$ that has two minima.
One, which is called the false vacuum (fv), with $V(\phi_{{\rm fv}})\equiv V_{{\rm fv}}$,
is positive and greater than the true vacuum (tv), $V(\phi_{{\rm tv}})\equiv V_{{\rm tv}},\,V_{{\rm tv}}<V_{{\rm fv}}$.
The problem is to calculate the transition rate from the false to
the true vacuum. In this work we will ignore the effects of gravity
and work in flat space with $G_{N}=0$, leaving the discussion of
gravitational theories to a separate paper.

Based on the corresponding quantum mechanics problem one would expect
the true ground state $|0>$ of the field in such a potential to be
a wave functional that is a superposition of the ground states in
a potential with just the one or the other minimum, with a much larger
overlap with tv than with fv. i.e. $|<\phi_{{\rm fv}}|0>|\ll|<\phi_{{\rm tv}}|0>|$.
The true ground state is certainly not an eigenstate of the field
operator with eigenvalue $\phi_{{\rm tv}}$ (with vacuum energy $V_{{\rm tv}})$,
both because of vacuum fluctuations and because the overlap with the
state $|\phi_{{\rm fv}}>$ though small, would certainly not be zero.

Such a calculation of the ground state wave function however does
not tell us what the transition rate would be to a final state\footnote{As discussed later, in a true decay we should locate $\phi_{{\rm tv}}$
in a runaway region of the potential, as in the right panel of figure
\eqref{fig:PotentialFlatSpace}.} which is localized around $\phi_{{\rm tv}}$, if say the initial
state is localized around $\phi_{{\rm fv}}$. This would require a
solution of the (functional) Schroedinger equation for a wave packet
functional of $\phi$ localized initially around $\phi_{{\rm fv}}$
and then following its time evolution.

Elaborating on the discussion in \citep{Cespedes:2020xpn}, we argue
that the state localized in fv should be treated as a resonance and
the decay rate (width of the resonance) should be calculated identifying
the relevant pole in the S-matrix. The lifetime of fv is then obtained
by constructing a wave packet initially localized at fv. This then
shows the relation to the starting point of Coleman's argument and
also why it is not necessary to make rather dubious analytic continuation
arguments to justify the tunneling formalism\footnote{Similar criticisms have been made in \citep{Andreassen:2016cvx} but
here we go further and in particular we explain in the Appendix why
we believe that the formalism developed there does not quite resolve
the problem.}. In particular there is no issue with negative modes, and consequently
the pre-factor is somewhat different with the time scale in the tunneling
formula set by the period of oscillations in the fv. Finally we address
the question of how to formulate properly, a WKB analysis of computing
the tunneling rate in a situation like that of the Higgs vacuum decay,
where the possibility of instability only arises after we include
quantum corrections, in contrast to the situation addressed by Coleman
and collaborators where the classical potential already has more than
one local minimum.

\section{The functional wave equation\label{sec:The-functional-wave}}

In this section we review the formalism developed in \citep{Cespedes:2020xpn},
based on earlier work by many authors\footnote{There is a long history for WKB beyond one particle quantum mechanics,
going back to the work of VanVleck\citep{VanVleck:1928zz}, see also
\citep{Brown:1971zzc} for a pedagogical discussion. More recent works
are \citep{Banks:1972xa,Banks:1973ps,Banks:1973uca}. Some relevant
works that we are familiar with are Bitar and Chang \citep{Bitar:1978vx}
and Gervais and Sakita \citep{Gervais:1977nv}, who first applied
this to field theory. Our discussion is closest to that found in \citep{Tanaka:1993ez}.}, and then explain some technical issues in more detail than was given
there\footnote{Although in this paper we confine ourselves to non-gravitational field
theory it is instructive to address the full formulation and then
specialize to the flat space case. }.

In the ADM formalism\footnote{For a review see for example \citep{Poisson:2009pwt}. },
the constraints of the system are\footnote{$N,N_{i}$ are the lapse and shift of the ADM formalism, $\pi_{N},\pi_{N_{i}}$
their conjugate momenta, and $\approx$ implies equality on a solution.} $\pi_{N}\approx0,\,\pi_{N_{i}}\approx0,$ the field spatial momentum
${\cal P}_{i}\approx0$ and the Hamiltonian\footnote{Although in this paper we only consider flat space tunneling, in this
section we address the general situation including gravity and specialize
to the flat space case later.} 
\begin{equation}
{\cal H=}\frac{1}{2}G^{MN}(\boldsymbol{\Phi})\pi_{M}\pi_{N}+f[\boldsymbol{\Phi}]\approx0.\label{eq:H}
\end{equation}
Here $\boldsymbol{\Phi}$ denotes the components of the spatial metric,
on some spatial slice of the foliation, as well as other fields and
their spatial derivatives (up to second order). The $\pi$'s are their
conjugate momenta. The metric on field space $G_{MN}$ is essentially
the metric on superspace with additional components coming from the
kinetic components of scalar fields etc. Up to operator ordering ambiguities
which are fixed by demanding the derivatives with respect to $\Phi^{M}$
are covariant with respect to the supermetric $G_{MN}$ we have the
Wheeler-DeWitt (WDW) eqn (replacing $\pi_{M}\rightarrow-i\hbar\nabla_{M}$)

\begin{equation}
{\cal H}\Psi=\left[-\frac{\hbar^{2}}{2}G^{MN}(\boldsymbol{\Phi})\nabla_{M}\nabla_{N}+f[\boldsymbol{\Phi}]\right]\Psi=0.\label{eq:WD}
\end{equation}

Note that the classical constraints $\pi_{N_{t}}\approx0,\,\pi_{N_{i}}\approx0$
imply that the wave function $\Psi$ is independent of $N,N_{i}$.
As usual we write, 
\begin{equation}
\Psi[\boldsymbol{\Phi}]=e^{\frac{i}{\hbar}S[\boldsymbol{\Phi}]}\label{eq:Psi}
\end{equation}
and define the semi-classical expansion 
\begin{equation}
S[\boldsymbol{\Phi}]=S_{0}[\boldsymbol{\Phi}]+\hbar S_{1}[\boldsymbol{\Phi}]+O(\hbar^{2})\label{eq:Sexpn}
\end{equation}
Substituting these two equations in the Wheeler DeWitt eqn we can
in principle determine recursively the semi-classical expansion coefficients.
The lowest two orders give, 
\begin{eqnarray}
\frac{1}{2}G^{MN}\frac{\delta S_{0}}{\delta\Phi^{M}}\frac{\delta S_{0}}{\delta\Phi^{N}}+f[\boldsymbol{\Phi}] & = & 0,\label{eq:HJ}\\
2G^{MN}\frac{\delta S_{0}}{\delta\Phi^{M}}\frac{\delta S_{1}}{\delta\Phi^{N}} & = & iG^{MN}\nabla_{M}\nabla_{N}S_{0}.\label{eq:prefactoreqn}
\end{eqnarray}

Observe that at a turning point $\pi_{M}=\frac{\delta S_{0}}{\delta\Phi^{M}}=0,\,\forall\,M,$
(which implies $f[\boldsymbol{\Phi}]=0$) the semi-classical expansion
breaks down since $S_{1}$ cannot be determined. Let us now introduce
a set of integral curves parametrized by $s$ on the field manifold
on the selected spatial slice defined by, 
\begin{equation}
C(s)\frac{d\Phi^{N}}{ds}=G^{MN}\frac{\delta S_{0}}{\delta\Phi^{M}}\label{eq:curves}
\end{equation}

Defining the metric on field space as 
\[
d\tau^{2}\equiv\int_{X}\delta\Phi^{M}G_{MN}\delta\Phi^{N},
\]
we get using \eqref{eq:curves} and \eqref{eq:HJ}, 
\begin{equation}
\left(\frac{d\tau}{ds}\right)^{2}=\int_{X}\frac{\delta\Phi^{M}}{ds}G_{MN}\frac{\delta\Phi^{N}}{ds}=-2C^{-2}(s)\int_{X}f[\Phi].\label{dtauds}
\end{equation}

The solution to the Hamilton-Jacobi eqn.\eqref{eq:HJ} and to the
leading quantum correction, may then be written as integrals over
the field space distance $\tau$ (i.e. putting $s=\tau$ in \eqref{dtauds}),
\begin{equation}
S_{0}[\Phi_{\tau}]=\int^{\tau}d\tau'\sqrt{-2\int_{X}f[\Phi_{\tau'}]},\label{eq:S0tau}
\end{equation}
and

\begin{align}
S_{1}[\boldsymbol{\Phi}_{\tau}] & =\frac{i}{2}\int^{\tau}d\tau'\frac{1}{\sqrt{-2\int_{X}f[\Phi_{\tau'}]}}\int_{X}\nabla^{2}S_{0}[\boldsymbol{\Phi}_{\tau'}]\label{eq:S1tau}\\
 & =\frac{i}{2}\ln\sqrt{-2\int_{X}f[\Phi_{\tau}]}+\frac{i}{2}\ln\sqrt{\left(\det G_{AB}\frac{d\Phi^{A}}{d\lambda^{\bar{N}}}\frac{d\Phi^{B}}{d\lambda^{\bar{M}}}\right)_{\tau}}+{\rm constant}.\label{eq:lndetterm}
\end{align}
Note that we have used $\tau$ to parametrize a trajectory in field
and $\lambda^{\bar{N}}$ to parametrize directions orthogonal to it
i..e $\delta\Phi^{M}(x)=\delta\tau t^{M}(x)+\delta\lambda^{\bar{P}}n_{\bar{P}}^{M}$,
with ${\bf t},{\bf n}_{\bar{P}}$ being the tangent and normal unit
vectors to the trajectory at $\boldsymbol{\Phi}$. Also we are using
DeWitt's condensed notation in that indices and summed (integrated)
over components of the field as well as spatial position. Finally
by definition the matrix in the second term in \eqref{eq:lndetterm}
does not have any zero eigenvalues since it represents a well-defined
coordinate transformation in field space whose metric is assumed to
be non-degenerate.

The semi-classical wave function is then \citep{Cespedes:2020xpn},

\begin{equation}
\Psi[\Phi_{\tau}]=\frac{\left[2\int_{X}f[\Phi_{0}]\right]^{1/4}}{\left[2\int_{X}f[\Phi_{\tau}]\right]^{1/4}}\frac{\left(\det G_{AB}\frac{d\Phi^{A}}{d\lambda^{\bar{N}}}\frac{d\Phi^{B}}{d\lambda^{\bar{M}}}\right)_{0}^{1/4}}{\left(\det G_{AB}\frac{d\Phi^{A}}{d\lambda^{\bar{N}}}\frac{d\Phi^{B}}{d\lambda^{\bar{M}}}\right)_{\tau}^{1/4}}e^{\frac{i}{\hbar}\left(S_{0}\left[\Phi_{\tau}\right]-S_{0}[\Phi_{0}]\right)}\Psi[\Phi_{0}],\label{eq:Psi1}
\end{equation}
with $S_{0}$ given by \eqref{eq:S0tau}. An alternative form is the
field theoretic version of the VanVleck formula, 
\begin{align}
\Psi[\Phi_{\tau}] & =\frac{\left[2\int_{X}f[\Phi_{0}]\right]^{1/4}}{\left[2\int_{X}f[\Phi_{\tau}]\right]^{1/4}}\frac{\left(\sqrt[4]{G_{0}}\right)}{\left(\sqrt[4]{G_{\tau}}\right)}\frac{\sqrt{\det\left[\frac{\delta^{2}|S_{0}|}{\delta\Phi^{A}\delta\alpha^{\bar{A}}}\right]_{\tau}}}{\sqrt{\det\left[\frac{\delta^{2}|S_{0}|}{\delta\Phi^{A}\delta\alpha^{\bar{A}}}\right]_{0}}}e^{\frac{i}{\hbar}S_{0}\left[\Phi\tau,\Phi_{0}\right]}\Psi[\Phi_{0}],\,\label{eq:VV}\\
S_{0}\left[\Phi_{\tau},\Phi_{0}\right] & =\int_{0}^{\tau}d\tau'\sqrt{-2\int_{X}f[\Phi_{\tau'}]}.\label{eq:S0}
\end{align}
Here the $\alpha$'s may be taken to be ``initial values'' of the
fields. This form of the pre-factor was first obtained by VanVleck
\citep{VanVleck:1928zz} (see also \citep{Brown:1971zzc}) for many
particle quantum mechanics. It is actually equal to the one loop contribution
to the effective action (see \citep{DeWitt-Morette:1976ydh}) so that
(with $K_{F}^{-1}$ being the Green's function with Feynman boundary
conditions), 
\begin{equation}
\Gamma_{1}\left[\Phi_{{\rm C}}(\phi_{\tau},\Phi_{0})\right]=-\frac{i}{2}\ln\det K_{F}^{-1}=-\frac{i}{2}\ln\det\left[\frac{\delta^{2}\vert S_{0}\left[\Phi_{\tau},\Phi_{0}\right]\vert}{\delta\Phi_{\tau}^{A}\delta\Phi_{0}^{\bar{A}}}\right],\label{eq:VVdet}
\end{equation}
is the one-loop contribution to the effective action (see eqns. (14.39),(21.97)
and (34.16) of DeWitt \citep{DeWitt:2003pm}), evaluated on a classical
solution $\Phi_{{\rm C}}$ with boundary values $\Phi_{\tau},\Phi_{0}$
and we've taken the arbitrary constants to be the values of the fields
at $\tau=0$. Hence the wave function may be written in the suggestive
form, 
\begin{equation}
\Psi[\Phi_{\tau}]=\frac{\left[2\int_{X}f[\Phi_{0}]\right]^{1/4}}{\left[2\int_{X}f[\Phi_{\tau}]\right]^{1/4}}\frac{\left(\sqrt[4]{G_{0}}\right)}{\left(\sqrt[4]{G_{\tau}}\right)}e^{\frac{i}{\hbar}\Gamma^{(1)}}\Psi[\Phi_{0}],\label{eq:Psi2-1}
\end{equation}
where $\Gamma^{(1)}\left[\Phi_{{\rm C}}(\phi_{\tau},\Phi_{0})\right]=S_{0}\left[\Phi_{\tau},\Phi_{0}\right]+\hbar\left(\Gamma_{1}\left[\Phi_{{\rm C}}(\phi_{\tau},\Phi_{0})\right]\right)-\Gamma_{1}\left[\Phi_{{\rm 0}}\right]$
is the quantum effective action\footnote{It is tempting to conjecture that the exact value of the exponent
is the exact quantum effective action evaluated at its extremum!} at one-loop order evaluated on the classical trajectory $\Phi_{C}$
minus that evaluated at the initial field configuration $\Phi_{0}$.

Finally we note that the classical path is the one that minimizes
the action functional - implying that the path satisfies the equation
of motion, 
\begin{equation}
\frac{D}{D\sigma}\frac{d\Phi_{\sigma}^{N}}{d\sigma}+G^{NP}\frac{\delta f[\Phi_{\sigma}]}{\delta\Phi^{P}}=0,\label{eq:eom}
\end{equation}
where the parameter $\sigma$ is related to $\tau$ the distance in
field space along the trajectory, by 
\begin{equation}
d\sigma=\frac{d\tau}{\sqrt{-2\int_{X}f[\Phi_{\tau}]}}.\label{eq:dsigmadtau}
\end{equation}
Note that this parametrization breaks down at the turning points (as
is the case with the WKB calculation of the wave function \eqref{eq:Psi2-1}).
Furthermore we emphasize that in gravitational theories the metric
on field space is not positive definite so that $d\tau^{2}$ may be
positive negative or zero. In the following we will focus on non-gravitational
field theories so that (at least in unitary gauge for a gauge theory)
$d\tau^{2}$ is positive definite. In this case $d\sigma$ may be
real or imaginary depending on the sign of $\int_{X}f[\Phi_{\tau}]$.
When the latter is positive (corresponding to the particle being under
the potential barrier in the QM case - see also the next section),
$d\sigma$ is pure imaginary and may be interpreted as Euclidean time.
However it should be stressed that this should not be viewed as analytically
continued real time since the equation we are solving is essentially
the time independent Schroedinger equation.

\section{Flat space tunneling of a scalar field - Coleman vs WKB\label{sec:Flat-space-tunneling}}

\begin{figure}[h!]
\centering{}\includegraphics[scale=0.7]{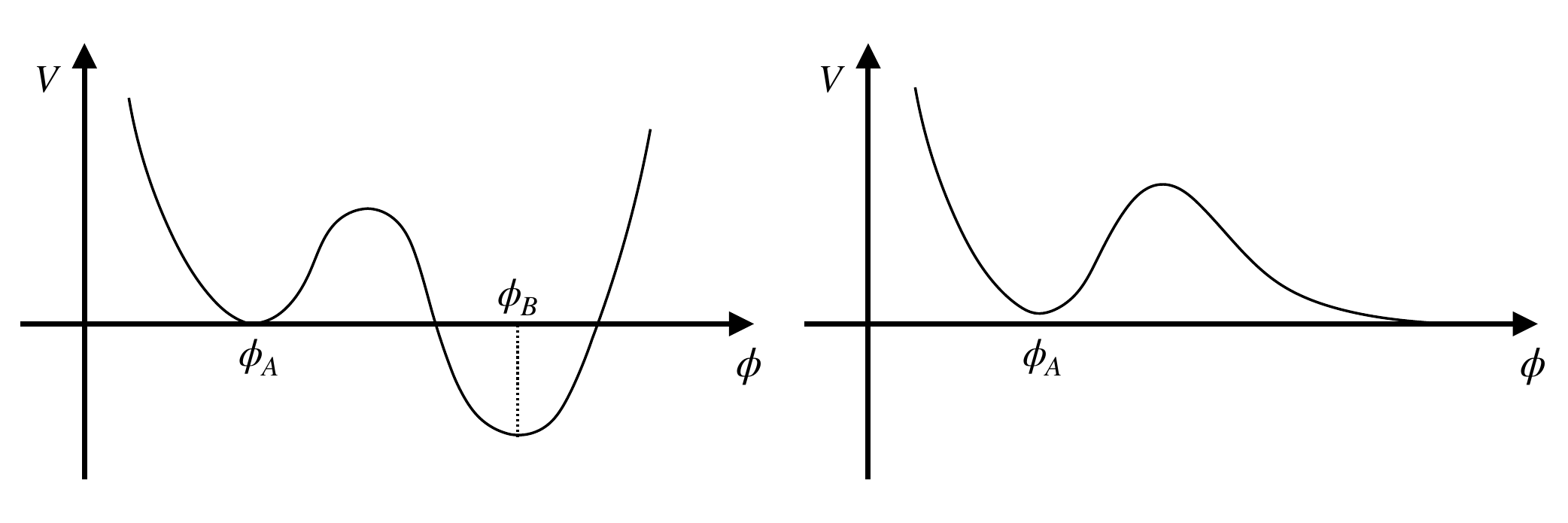} \caption{Potential V{[}$\phi${]}.\label{fig:PotentialFlatSpace}}
\end{figure}

\begin{figure}[h!]
\centering{}\includegraphics[scale=0.7]{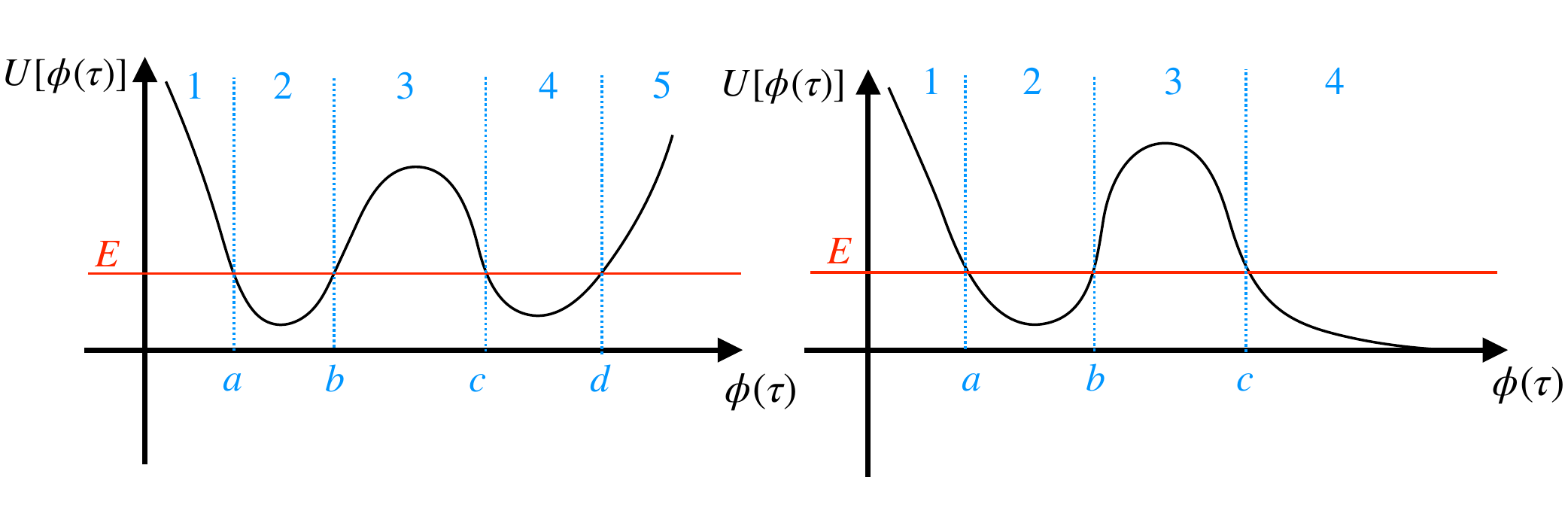} \caption{Potential U{[}$\phi(\tau)${]}.\label{fig:PotentialUFlatSpace}}
\end{figure}

Consider the potentials of figure~(\ref{fig:PotentialFlatSpace}).
We need to solve the flat space version of eqn. \eqref{eq:WD}. Now
we have a global constraint for a state with a given energy - classically
it is a constraint on the Hamiltonian (rather than the density), 
\begin{equation}
\hat{H}=\int_{X}\left[\frac{\pi_{\phi}^{2}}{2}+\frac{1}{2}\left(\boldsymbol{\nabla}_{x}\phi\right)^{2}+V(\phi)\right]=E\label{eq:Hphi}
\end{equation}
and hence we get the Schroedinger equation 
\begin{equation}
\int_{X}\left[-\frac{\hbar^{2}}{2}\frac{\delta^{2}}{\delta\phi(x)^{2}}+\frac{1}{2}\left(\boldsymbol{\nabla}_{x}\phi\right)^{2}+V(\phi)\right]\Psi[\phi]=E\Psi[\phi].\label{eq:S-eqn}
\end{equation}
The effective quantum mechanical potential is $U\left[\phi\right]=\int_{X}\left[\frac{1}{2}\left(\boldsymbol{\nabla}_{x}\phi\right)^{2}+V(\phi)\right]$.
This is shown in Fig.~\ref{fig:PotentialUFlatSpace} as a function
of a parametrized path $\phi=\phi(\tau)$ in field space. 

If the space $X$ is not compact, in order to have these integrals
well-defined we need to consider field configurations with compact
support - say with a size $R$. The semi-classical equations \eqref{eq:HJ}\eqref{eq:prefactoreqn}
now have $G^{\phi\phi}=1$. and $\int_{X}f\left[\phi\right]=\int_{x}\left[V\left(\phi\right)\right]-E$.
Substituting this into \eqref{eq:S0tau}\ref{eq:S1tau} gives the
corresponding expressions for $S_{0},S_{1}$.

Let us write 
\begin{align}
k\left(\tau\right) & =k_{0}+\hbar k_{1},\,k_{0}=\sqrt{2\left[E-U[\phi_{\tau}]\right]},\,k_{1}=\frac{1}{2}\frac{i}{\sqrt{2\left[E-U[\phi_{\tau}]\right]}}\int_{X}\nabla^{2}S_{0}[\boldsymbol{\phi}_{\tau}]\,{\rm for}\,E>U\left(\phi_{\tau}\right).\label{eq:ktau}\\
\kappa\left(\tau\right) & =\kappa_{0}+\hbar\kappa_{1},\,\kappa_{0}=\sqrt{2\left[U[\phi_{\tau}]-E\right]},\,\kappa_{1}=\frac{1}{2}\frac{i}{\sqrt{2\left[U[\phi_{\tau}]-E\right]}}\int_{X}\nabla^{2}S_{0}[\boldsymbol{\phi}_{\tau}],\,{\rm for}\,E<U(\phi_{\tau}).\label{eq:kappatau}
\end{align}
Note that both $k_{0}$ and $\kappa_{0}$ as well as $ik_{1}$ and
$\kappa_{1}$ are real (recall that for $E>U$ $S_{0}$ is real while
for $E<U$ $S_{0}$ is pure imaginary). We then have for the wave
functions in the four regions of figure \eqref{fig:PotentialUFlatSpace}
\begin{equation}
\Psi[\phi_{\tau}]=\begin{cases}
A_{I}\frac{1}{\sqrt{\kappa_{0}\left(\tau\right)}}\exp\left\{ -\int_{\tau}^{a}\tilde{\kappa}\left(\tau\right)d\tau'\right\}  & +B_{I}\frac{1}{\sqrt{\kappa_{0}\left(\tau\right)}}\exp\left\{ +\int_{\tau}^{a}\tilde{\kappa}\left(\tau\right)d\tau'\right\} \\
A_{II}\frac{1}{\sqrt{k_{0}\left(\tau\right)}}\exp\left\{ i\int_{a}^{\tau}\tilde{k}\left(\tau'\right)d\tau'\right\}  & +B_{II}\frac{1}{\sqrt{k_{0}\left(\tau\right)}}\exp\left\{ -i\int_{a}^{\tau}\tilde{k}\left(\tau'\right)d\tau'\right\} \\
A_{III}\frac{1}{\sqrt{\kappa_{0}\left(\tau\right)}}\exp\left\{ -\int_{b}^{\tau}\tilde{\kappa}\left(\tau'\right)d\tau'\right\}  & +B_{III}\frac{1}{\sqrt{\kappa_{0}\left(\tau\right)}}\exp\left\{ -\int_{b}^{\tau}\tilde{\kappa}\left(\tau\right)d\tau'\right\} \\
A_{IV}\frac{1}{\sqrt{k_{0}\left(\tau\right)}}\exp\left\{ i\int_{c}^{\tau}\tilde{k}\left(\tau'\right)d\tau'\right\}  & +B_{IV}\frac{1}{\sqrt{k_{0}\left(\tau\right)}}\exp\left\{ -i\int_{c}^{\tau}\tilde{k}\left(\tau'\right)d\tau'\right\} 
\end{cases}\label{eq:Psiphi}
\end{equation}
In the above we've redefined $k$ and $\kappa$ so that we may use
the standard WKB matching conditions, which means (from eqn. \eqref{eq:lndetterm}),
\begin{equation}
\tilde{\kappa}\left(\tau\right)\equiv\kappa_{0}-\frac{1}{2}\partial_{\tau}\ln\det\frac{d\phi(\tau)}{d\lambda^{\bar{N}}},\,\tilde{k}\left(\tau\right)\equiv k_{0}+\frac{i}{2}\partial_{\tau}\ln\det\frac{d\phi(\tau)}{d\lambda^{\bar{N}}}\label{eq:kappaktilde}
\end{equation}

The coefficients $A_{I}\ldots B_{IV}$ in \eqref{eq:Psiphi} are related
by the WKB matching conditions if one ignored the transverse fluctuations
in the second terms on the RHS of the equations above. We will however
assume that this matching is still valid even in the presence of these
terms. So under this assumption, i.e. using the WKB matching conditions
of QM, we get 
\begin{equation}
\begin{bmatrix}A_{II}\\
B_{II}
\end{bmatrix}=\begin{bmatrix}e^{-i\pi/4} & -e^{-i\pi/4}/2i\\
e^{i\pi/4} & e^{i\pi/4}/2i
\end{bmatrix}\begin{bmatrix}A_{I}\\
B_{I}
\end{bmatrix}.\label{eq:ItoII}
\end{equation}
The coefficients $A_{IV},B_{IV}$ are related to $A_{II},B_{II}$
by the standard formulae as given for instance in Merzbacher \citep{Merzbacher:1998}
eqn. (7.47), provided we identify $F,G$ and $A,B$ therein respectively
with $A_{IV},B_{IV}$ and $e^{iL}A_{II},e^{-iL}B_{II}$, where $L=\int_{a}^{b}k_{\tau'}d\tau$.
So 
\[
\begin{bmatrix}A_{IV}\\
B_{IV}
\end{bmatrix}=\frac{1}{2}\begin{bmatrix}1/2\theta+2\theta & i\left(1/2\theta-2\theta\right)\\
-i\left(1/2\theta-2\theta\right) & 1/2\theta+2\theta
\end{bmatrix}\begin{bmatrix}e^{iL}A_{II}\\
e^{-iL}B_{II}
\end{bmatrix}.
\]
Here 
\begin{equation}
\theta\equiv\exp\left[\int_{b}^{c}\tilde{\kappa}\left(\tau\right)d\tau\right]=\sqrt{D(\tau_{c},\tau_{b})}e^{\int_{b}^{c}d\tau\sqrt{2\left[U[\phi_{\tau}]-E\right]}},\label{eq:theta}
\end{equation}
where $D$ is the VanVleck-Morette determinant, 
\begin{equation}
D(\tau_{c},\tau_{b})=\det\left[\frac{\delta^{2}\vert S_{0}\left[\phi_{c},\phi_{b}\right]\vert}{\delta\phi_{c}\delta\phi_{b}}\right].\label{eq:D}
\end{equation}
Finally we need to use the condition that on the left hand side the
potential rises indefinitely so that the wave function in region $I$
must decrease to the left, i.e. we need to set $B_{I}=0$. Thus we
get 
\begin{align}
A_{IV} & =e^{-i\pi/4}\left(2\theta\cos L+i\frac{1}{2\theta}\sin L\right)A_{I}\label{eq:A4}\\
B_{!V} & =e^{i\pi/4}\left(2\theta\cos L-i\frac{1}{2\theta}\sin L\right)A_{I}\label{eq:B4}
\end{align}
What we have is thus a state which comes in from the right, tunnels
through the barrier, is reflected off the wall on the LHS, and tunnels
back through the barrier, to give an outgoing state to the right.
The S-matrix is a phase ($S=A_{IV}/B_{IV}$) with complex poles corresponding
to the bound states/resonances in region II. To identify the decay
widths of the corresponding resonances we first identify the (putative)
bound states (which would have existed if the barrier were impenetrable)
- these correspond to 
\[
\cos L=0\Rightarrow L=\left(n+\frac{1}{2}\right)\pi,{\rm \,\,equivalently}E=E_{n}\,({\rm say}),
\]
where $n=0$ corresponds to the ground state. Expanding around this
state, we write (for a discussion of a related problem in QM see \citep{Merzbacher:1998}
chapter 7 section 4), 
\begin{equation}
\cos L\simeq\mp(E-E_{0})\left(\frac{\partial L}{\partial E}\right)_{0},\sin L\simeq1,\label{eq:cosL}
\end{equation}
and get (see for example \citep{Merzbacher:1998} eqn (13.87) 
\[
S\equiv\frac{A_{IV}}{B_{IV}}=\frac{E-E_{0}-i\left\{ 1/\left(4\theta{}^{2}\left(\frac{\partial L}{\partial E}\right)_{0}\right)\right\} }{E-E_{0}+i\left\{ 1/(\left(4\theta{}^{2}\left(\frac{\partial L}{\partial E}\right)_{0}\right)\right\} }\equiv e^{2i\phi}=\frac{E-E_{0}-i\Gamma_{0}/2}{E-E_{0}+i\Gamma_{0}/2}.
\]
The decay width is given by 
\begin{equation}
\Gamma_{0}=\frac{1}{2\theta^{2}}\left(\frac{\partial L}{\partial E}\right)_{0}^{-1}.\label{eq:Gamma}
\end{equation}
The phase shift $\phi$ has the standard form 
\[
\tan\phi=\frac{\Gamma_{0}/2}{E-E_{0}},
\]
with the lifetime of the resonance given by, 
\begin{align}
\tau & =\Gamma_{0}^{-1}=2\theta^{2}\left(\frac{\partial L}{\partial E}\right)_{0}=\left[\frac{\partial}{\partial E}\left\{ \int_{a}^{b}d\tau\sqrt{2\left(E-U(\tau\right)}\right\} \right]\vert_{E=E_{0}}2D(\tau_{c},\tau_{b})\exp\left[2\int_{b}^{c}d\tau\sqrt{2\left(U(\tau\right)-E_{0}})\right].\label{eq:lifetime}\\
U(\tau) & =\int_{X}\left(\frac{1}{2}\left(\boldsymbol{\nabla}_{x}\phi_{\tau}\right)^{2}+V(\phi_{\tau})\right)\nonumber 
\end{align}
Note that we've ignored a quantum correction due to the transverse
fluctuations $D(a,b)$ in the expression for $L$, so $2\left(\frac{\partial L}{\partial E}\right)_{0}$
is the classical period for oscillations between $a$ to $b$ and
one could have divided the tunneling probability i.e. the WKB factor
$1/\theta^{2}=\frac{1}{D(b,c)}\exp\left[-2\int_{b}^{c}d\tau\sqrt{2\left(U(\tau\right)-E})\right]$,
by this to get the first factor of \eqref{eq:lifetime} formula heuristically.
Here we've derived it purely from quantum mechanical considerations.
In the next subsection we'll review the justification for identifying
$\Gamma^{-1}$ as the life time of the resonance.

Finally we note that the leading contribution to the lifetime/rate
namely the argument of the exponential in eqn. \eqref{eq:lifetime}
can be rewritten as a classical Euclidean action \citep{Coleman:1977py},
if we change the integration variable from $\tau$ to $\sigma$ defined
by eqn. \eqref{eq:dsigmadtau}. In fact one may rewrite the exponent,
noting the symmetry under $\tau\rightarrow-\tau$ and using $2\int_{b}^{c}=\oint_{b}$
i.e. an integral from $b$ to $c$ and back again, and observing that
$\tau\rightarrow b$ corresponds to $\sigma\rightarrow\pm\infty$,
as $B\equiv\int_{-\infty}^{+\infty}2\left(U(\tau\left(\sigma\right)-E\right)$.
Also the energy conservation constraint \eqref{eq:Hphi} after replacing
$\pi^{2}=\dot{\phi^{2}}\rightarrow-\left(\frac{d\phi}{d\sigma}\right)^{2}$,
gives $S_{E}\equiv\int_{-\infty}^{+\infty}d\sigma\int_{X}\frac{1}{2}\left(\frac{d\phi}{d\sigma}\right)^{2}+U=\int_{-\infty}^{+\infty}d\sigma\left[2\left(U-E\right)+E\right]$.
But we also have $\int_{-\infty}^{+\infty}d\sigma E=S_{E}^{{\rm fv}}$
the action at the false vacuum. Furthermore the configuration which
minimizes $S_{E}$ is the $O(4)$ symmetric bounce \citep{Coleman:1977th}
corresponding to a solution $\phi(r),\,r^{2}\equiv{\bf \tau^{2}+|{\bf x}|^{2}}$
with boundary conditions $\phi(\infty)=\phi_{II},\,\frac{d\phi}{dr}\vert_{r=0}=0$
where $\phi_{II}$ is the field configuration in the classically allowed
false vacuum well region $II$ in the figure. Hence one gets Coleman's
formula $B=S_{E}^{{\rm b}}-S_{E}^{{\rm fv}}$. Thus our final result
for the lifetime of the resonance \eqref{eq:lifetime} is, 
\begin{equation}
T=\left[\frac{\partial}{\partial E}\left\{ \int_{a}^{b}d\tau\sqrt{2\left(E-U(\tau\right)}\right\} \right]\vert_{E=E_{0}}2D(\tau_{c},\tau_{b})\exp\left[S_{E}^{{\rm b}}-S_{E}^{{\rm fv}}\right].\label{eq:lifetime2}
\end{equation}

This derivation did not involve either the dilute gas approximation
or an analytic continuation (which as we argue in the appendix is
anyway rather dubious). We have rigorously followed\footnote{Except for the assumption regarding the matching conditions - see
discussion after eqn. \eqref{eq:kappaktilde}.} the standard discussion of the WKB expansion in quantum mechanics
and generalized it to field theory. Our formula agrees with Coleman's
as far as the leading (exponential) term is concerned. However the
first quantum correction (i.e. the pre-factor) is somewhat different.

Firstly Coleman's formula had two infinite factors, $TV$ the total
time and volume of space time, and dividing by these gave him a formula
for the decay rate per unit volume per unit time. Here we do not have
these factors - instead the time scale is set by the sloshing time
for the field configuration which is of the order of the inverse of
a characteristic mass scale in region $II$. Also the size of the
decaying configuration is determined by the size \textbf{$R_{{\rm b}}$}
of the nucleated bubble which in turn is obtained by finding the bounce
solution which minimizes the action. Consequently the decay rate per
unit volume is obtained by dividing the rate in (\ref{eq:Gamma})
by $R_{{\rm b}}^{3}$ and hence the total decay rate of the universe
would be obtained by multiplying this by the Hubble volume.

In the example of the Higgs vacuum decay (assuming that the classical
action can be replaced by the effective action see discussion in section
\eqref{sec:Decay-of-the}), the sloshing time scale is set by the
physical Higgs mass so we would expect the factor in square brackets
in \eqref{eq:lifetime2} to be $O(m_{{\rm Higgs}}^{-1})$. Hence the
formula for the lifetime of a Hubble volume ($R_{0}^{3}=H_{0}^{-3})$
of the standard model universe takes the form 
\begin{equation}
T\simeq m_{{\rm Higgs}}^{-1}\frac{R_{{\rm b}}^{3}}{R_{0}^{3}}2D(\tau_{c},\tau_{b})\exp\left[S_{E}^{{\rm b}}-S_{E}^{{\rm fv}}\right]\label{eq:lifetimeSM}
\end{equation}

The prefactor however has the usual divergence difficulties of one-loop
perturbation theory, since the third factor is the (exponential of)
the one loop term in the quantum effective action evaluated on the
bounce (see eqn.\eqref{eq:VVdet}). One needs to regularize and renormalize
it in order to have a meaningful result. Let us choose to do this
using the proper time representation of the (log of the) determinant
of $K$. So we have\footnote{We keep a general metric here for clarity though most of the discussion
in this work will be in flat space.} 
\begin{equation}
{\rm Tr}\ln K_{F}[\Phi_{C})=\int d^{4}x\sqrt{g}\int_{1/\Lambda^{2}}^{\infty}\frac{ds}{s}{\rm tr}e^{is\left(K\left[\Phi_{C}\right]+i\epsilon\right)}\label{eq:Trln}
\end{equation}
The terms which diverge for large $\Lambda$ would need to be subtracted
and the effective action renormalized at the scale set by the size
of the bounce. We also observe that the leading (field independent)
terms in the above just go towards renormalizing the cosmological
constant. In particular the zero modes associated with the bounce
(for instance Coleman's solution) is a logarithmic renormalization
of the cosmological constant (the minimum of the potential energy)
due to vacuum fluctuations in the background of the bounce, that is
proportional to the dimension of the zero-mode space.

\subsection{Wave packet solution\label{subsec:Wave-packet-solution}}

The discussion in the previous section parallels the corresponding
one in quantum mechanics. The system has a fixed energy and the physics
is described by the time independent Schroedinger equation. To actually
discuss the time evolution of a state one needs to construct a wave
packet. This will then justify the relation between $\Gamma$ and
the time rate of decay of a resonance provided certain conditions
are satisfied. For simplicity we'll take one dimensional quantum mechanics\footnote{A similar discussion for the particular case of a square potential
can be found in \citep{Andreassen:2016cvx}.}. While the previous considerations involved an energy eigenstate,
to get time dependence we need to take a superposition of energy eigenstates.

Let us now consider a wave packet $\psi(x,t)$ with $\psi(x,0)$ localized
in region II of figure \eqref{fig:PotentialUFlatSpace}. Expanding
in energy eigenstates $u_{E}(x)$ we have 
\begin{equation}
\psi(x,t)=\int dy\psi(y,0)\int dEe^{-iEt}u_{E}(x)u_{E}^{*}(y).\label{eq:psipsi0}
\end{equation}
Now from the relations in the previous subsection (choosing without
loss of generality $A_{I}$ to be real so $B_{II}=A_{II}^{*}$ and
$B_{IV}=A_{IV}^{*}$), 
\begin{equation}
A_{II}=\frac{2\theta^{-1}\left(\frac{\partial L}{\partial E}\right)^{-1}}{E-\left(E_{0}-i\Gamma_{0}/2\right)}A_{IV}.\label{eq:AIIIV}
\end{equation}
We used here equations \eqref{eq:AIIIV}\eqref{eq:cosL}\eqref{eq:Gamma}.
Thus we expect the energy eigenfunctions to have poles corresponding
to a resonances at $E=E_{0(n)}-i\Gamma_{0(n)}$. So we may write (just
focussing on the lowest energy state) $u_{E}(x)=f_{E}(x)/\left(E-\left(E_{0}-i\Gamma_{0}/2\right)\right)$,
where $f_{E}(x)$ is non-singular at $E=E_{c}\equiv E_{0}-i\Gamma_{0}/2$.
Thus \eqref{eq:psipsi0} becomes,
\begin{align}
\psi(x,t) & =\int dy\psi(y,0)\int_{0}^{\infty}dEe^{-iEt}\frac{f_{E}(x)f_{E}^{*}(y)}{\left(E-E_{c}\right)\left(E-E_{c}^{*}\right)}\nonumber \\
 & =\int dy\psi(y,0)\left\{ \left(\int_{C}-\int_{-i\infty}^{0-i\epsilon}-\int_{\Gamma}\right)dEe^{-iEt}\frac{f_{E}(x)f_{E}^{*}(y)}{\left(E-Ec\right)\left(E-E_{c}^{*}\right)}\right\} ,\label{eq:psiintegral}
\end{align}
where we've taken the initial wave function $\psi(y,0)$ to have significant
overlap only with the ground state wave function. In the second line
the integral over real positive $E$ was rewritten as that over a
closed contour $C$ from$-i\infty$ to $0$, $0$ to infinity and
an arc $\Gamma$ in the south-east quadrant from $0_{-}$ to $-\pi/2$
. The integral over the arc vanishes in the limit as $|E|\rightarrow\infty$
since $t$ is real and positive. Replacing the integration variable
in the second integral by putting $E\rightarrow-i{\cal E}$ and picking
up the residue of the pole at $E=E_{c}$, we may now write the wave
function as 
\begin{align}
\psi(x,t) & =\frac{2\pi}{\Gamma}f_{E_{c}}(x)\left\{ \int dy\psi(y,0)f_{E_{c}}^{*}(y)\right\} e^{-\frac{\Gamma}{2}t}e^{-iE_{0}t}\nonumber \\
 & +\int dy\psi(y,0)\int_{0}^{\infty}d{\cal E}\frac{f_{{\cal E}}(x)f_{{\cal E}}^{*}(y)}{\left(-i{\cal E}-E_{c}^{*}\right)\left(-i{\cal E}-E_{c}\right)}e^{-{\cal E}t}.\label{eq:psiintegral2}
\end{align}
Now since $\psi(y)$ has significant overlap with $f_{{\cal E}}$
only in an interval $(E_{0}-\Delta E,E_{0}+\Delta E)$ the second
term on the RHS may be rewritten as 
\[
\simeq\int dy\psi(y,0)\int_{E_{0}-\Delta E}^{E_{o}+\Delta E}d{\cal E}\frac{f_{{\cal E}}(x)f_{{\cal E}}^{*}(y)}{\left(-i{\cal E}-E_{c}^{*}\right)\left(-i{\cal E}-E_{c}\right)}e^{-E_{0}t}.
\]
Also assuming that the width of the resonances is smaller than the
energy i.e. $\Gamma\ll E_{0}$, for late times $1\ll\frac{1}{2}\Gamma t\ll E_{0}t$
we see that the second term in \eqref{eq:psiintegral2} is exponentially
suppressed compared to the first. Thus we have 
\begin{equation}
\psi(x,t)\simeq\frac{2\pi}{\Gamma}f_{E_{c}}(x)\left\{ \int dy\psi(y,0)f_{E_{c}}^{*}(y)\right\} e^{-\frac{\Gamma_{0}}{2}t}e^{-iE_{0}t},\,\,t\gg2\Gamma_{0}^{-1}\gg E_{0}^{-1}\label{eq:psidecay}
\end{equation}
and the probability of being in the ``false vacuum'' is $P_{{\rm fv}}\simeq\int_{II}dx|\psi|^{2}$.
Hence, as expected, the decay rate is 
\[
-\frac{1}{P_{{\rm fv}}}\frac{dP_{{\rm fv}}}{dt}=\Gamma_{0}.
\]
This also justifies the usual interpretation\eqref{eq:lifetime} of
$\Gamma^{-1}\equiv\tau$ as the lifetime of the unstable (resonance)
state. Note that if $\psi(y,0)$ has overlap with more than one resonance
then the RHS of \eqref{eq:psidecay} will involve a sum over all them.
In this case long time effects may come in if for some higher resonance
$n$, $\Gamma_{n}\ll\Gamma_{0}$, so that even if $\int\psi(y,0)f_{E_{n}}^{*}(y)dy\ll1$
this term will become dominant.

It is often stated that an unstable state can be regarded as one with
a complex energy. This statement is based on the above approximate
formula \eqref{eq:psidecay} which appears to imply that the wave
function is an eigenstate of energy with a complex eigenvalue. However
this is misleading since this formula (as is seen from its derivation),
is approximate, and therefore such a conclusion is unwarranted. Nevertheless
Coleman's argument is based on manipulating a functional integral
to find such a complex energy eigenvalue.

The argument in this subsection can clearly be generalized to QFT
both in a flat and a fixed (non-dynamical) curved background. However
it is unclear how to deal with a situation in which gravity is dynamical,
since in that case we are confronted with the well-known problem of
time and indeed the construction of a wave packet is problematic since
the WdW equation does not appear to permit it.

\subsection{Comparison to Coleman's formula\label{subsec:Comparison-to-Coleman's}}

How is this related to the well-known formula derived by Coleman \citep{Coleman:1977py,Callan:1977pt}? 
\begin{itemize}
\item The Hamiltonian ${\cal H}$ is Hermitian - it cannot have complex
eigenvalues. Coleman makes essential use of the heuristic interpretation
of $\Gamma$ as the imaginary part of an eigenvalue of ${\cal H}$
and the dilute gas approximation in order to get the decay rate. The
above calculation uses no such interpretation - $\Gamma$ is extracted
from the $S$-matrix for the scattering of a field configuration from
a potential which can accommodate quasi-bound states (resonances). 
\item The Coleman calculation disagrees in the pre-factor from the standard
WKB calculation (i.e. essentially eqn \eqref{eq:lifetime} with $U$
now being the quantum mechanical potential. In particular the Coleman
formula for the decay rate $\Gamma$ involves dividing by the (infinite)
translation symmetry of his instanton. i.e. the prefactor for the
transition probability contains a factor $T\rightarrow\infty$ that
is divided out to get the rate. In the calculation above there is
no such factor and effectively the division is by the classical period
for oscillations in the potential well. 
\item The last factor of \eqref{eq:lifetime} is of course the square of
the standard WKB decay amplitude and naturally is the same as in Coleman's
formula, with the identification of the exponent as the difference
between the Euclidean classical action for a classical solution with
the appropriate boundary conditions, and the action for a particle
whose position is localized in the well (at its minimum). 
\item In the field theory case Coleman and collaborators gave an argument
for the dominant contribution to this last factor to come from a Euclidean
4-sphere configuration \citep{Coleman:1977th}. However in the actual
evaluation of the decay amplitude the thin wall approximation is used
in which the under the barrier region effectively shrinks to a brane.
On the other hand propagation in the classically allowed region to
the right of the barrier is described in terms of the analytic continuation
of this Euclidean instanton. The logic of this situation seems somewhat
unclear. 
\item The Coleman argument for the decay width comes from having one (and
only one) negative eigenvalue for the fluctuation matrix around the
so-called bounce solution. This is then supposed to give an imaginary
part to what is manifestly a real integral. This does not make any
sense to us and arises because the integral $\int_{0}^{\infty}e^{\lambda x^{2}}dx,\,\lambda>0$
which is obviously divergent, has been defined to be that obtained
from $\int_{0}^{\infty}e^{-\lambda x^{2}}dx=\frac{1}{2}\sqrt{\frac{\pi}{\lambda}},\,\lambda>0$
by replacing $\lambda$ by $-\lambda$. It's not clear that this is
justified (see the Appendix where we discuss a recent attempt\citep{Andreassen:2016cvx}
to justify Coleman's calculation). By contrast our modification of
Coleman's argument obtained by generalization of WKB quantum mechanics
does not rely on having a single negative eigenvalue. 
\end{itemize}

\section{Decay of the Higgs vacuum\label{sec:Decay-of-the} }

In the Schroedinger equation \eqref{eq:S-eqn}, both the field space
metric $G_{MN}$ and the potential $f[\Phi]$ were taken to have their
classical values. However although in QM this is sufficient (since
the potential for example is external to the system), in QFT the potential
is a function of the dynamical variables (fields), and gets corrected
by quantum effects, so that one should replace it (as well as the
metric) by its quantum corrected value. This means that the WKB prefactor
would not be the entire $O(\hbar)$ correction. This is particularly
important in the case of the standard model where the quantum corrections
to the Higgs potential can change its qualitative features, leading
to an instability at large values of the Higgs field\footnote{For a recent review see \citep{Markkanen:2018pdo}.}.

In fact what is often done in the literature is to find a bounce solution
to the tunneling dynamics by using the effective potential $V_{{\rm eff}}=-\frac{1}{4}|\lambda_{{\rm eff}}|\phi^{4}$
where $\phi$ is the Higgs field and treating this as a classical
potential and $\lambda_{{\rm eff}}$ is an effective coupling at large
field values. However this is certainly not the classical Higgs potential
which (ignoring the mass term since it plays no role given that the
relevant scales are much larger than the EW scale) is $V_{{\rm classical}}=\frac{1}{4}\lambda_{0}\phi^{4}$
where $\lambda_{0}\simeq0.13$. In fact this sets the initial value
(along with the electro-weak value for the top coupling $y_{T}$,
etc) for integrating the RG equations, which are a complicated set
of coupled non-linear differential equations for the couplings $\lambda,y_{T}$,
etc. The tunneling action coming from the instanton is then proportional
to $1/|\lambda_{{\rm eff}}|$ and is minimized at the point where
$|\lambda_{{\rm eff}}|$ is a maximum corresponding to a zero of the
beta function. Clearly applying the standard WKB analysis (discussed
in the previous sections) is not straightforward.

Let us consider the one-loop corrected effective potential for the
standard model. For simplicity we will just consider the Higgs and
the top quark contributions to it (i.e. ignoring the Higgs mass term,
gauge couplings and light fermions), 
\begin{align}
V_{{\rm eff}} & =+\frac{1}{4}\lambda\left(\mu_{0}\right)\phi^{4}+\frac{\hbar}{64\pi^{2}}\left[M^{4}\left(\phi\right)\left(\ln\frac{M^{2}\left(\phi\right)}{\mu_{0}^{2}}-\frac{3}{2}\right)-12m_{{\rm t}}^{4}\left(\phi\right)\left(\ln\frac{m_{{\rm {\rm t}}}^{2}\left(\phi\right)}{\mu_{0}^{2}}-\frac{3}{2}\right)\right]\equiv\frac{1}{4}\lambda_{{\rm eff}}\left(\phi\right)\phi^{4},\label{eq:Veff1}\\
M^{2} & =3\lambda\phi^{2},\,\,m_{{\rm t}}^{2}=\frac{1}{2}y^{2}\phi^{2},\label{eq:Mandm}
\end{align}
where the top Yukawa coupling has been defined as $\frac{1}{\sqrt{2}}y_{{\rm }}\phi\bar{\psi}_{{\rm t}}\psi_{{\rm t}}$.
Here we may take the renormalization scale $\mu_{0}=m_{{\rm t}}^{{\rm phys}}\simeq173GeV$,
to be the the physical top mass (as a representative Electro-Weak
scale), and consider the first term in to be the classical potential.
In the WKB framework this is what goes into the classical action $S_{0}$
of sections \eqref{sec:The-functional-wave}\eqref{sec:Flat-space-tunneling}.
However this classical potential does not admit tunneling since the
standard model minimum is protected by an infinite barrier!

Nevertheless (as was originally observed long ago \citep{Krive:1976sg})
due to the fermion (in particular the top quark) contribution, the
one loop term (depending on the parameters), can dominate the positive
contribution, giving a situation where $\lambda_{{\rm eff}}(\phi)$
can become negative and making the standard model unstable or metastable.
In particular we see that for large $\phi$, 
\begin{equation}
\lambda_{{\rm eff}}\left(\phi\right)=\lambda_{0}+\frac{\hbar}{32\pi^{2}}\left[9\lambda_{0}^{2}-12y_{0}^{4}\right]\left(\ln\frac{\phi}{\mu_{0}}\right).\label{eq:lambda_phi_large}
\end{equation}
Identifying the classical couplings with the physical values from
the measured Higgs mass $125GeV$ and top mass $173GeV=\mu_{0}$,
gives $\lambda_{0}\simeq0.13$ and $y_{0}\simeq0.98$ so the second
term on the RHS is negative for large $\phi$. So the effective potential
goes through zero at a scale $\phi_{0}=\mu_{0}\exp\left[32\pi^{2}\lambda/\left(12y_{0}^{2}-9\lambda_{0}^{2}\right)\right]\sim6\times10^{3}GeV$.
However this value is clearly exponentially sensitive to the input
value of $y_{{\rm t}}$. In fact the numerical solution to the beta
function equations (with initial values for the couplings taken at
as the measured ones i.e. at the EW scale), gives $\lambda(\mu)=0,$
at $\mu\sim10^{9}GeV$. The point is that the top Yukawa coupling
decreases with $\mu$ as a consequence of the effect of the QCD coupling
$g\equiv g_{3}$, and so the RG improved equations at large scales
give significantly different results\footnote{See for example \citep{Degrassi:2012ry}. For sensitivity to Planck
scale corrections see \citep{Branchina:2014rva}. } from the above naive one-loop calculation.

\subsection{WKB for a quantum corrected action\label{subsec:WKB-for-a}}

Even though the one-loop calculation of the Higgs instability and
decay rate are rather different from the result obtained from integrating
the beta-functions, it provides us with a simpler situation in which
to revisit the argument that led to eqn. \eqref{eq:VV}. In other
words our point is that this allows us to apply the WKB method to
compute tunneling at least to one-loop order.

We replace \eqref{eq:WD} by the following quantum corrected equation,
\begin{equation}
\hat{H}\Psi[\Phi]=\left[-\frac{\hbar^{2}}{2}\left\{ G_{0}^{MN}(\boldsymbol{\Phi})+\hbar G_{1}^{MN}(\boldsymbol{\Phi})+O\left(\hbar^{2}\right)\right\} \nabla_{M}\nabla_{N}+\left\{ f_{0}[\boldsymbol{\Phi}]+\hbar f_{1}[\boldsymbol{\Phi}]+O\left(\hbar^{2}\right)\right\} \right]\Psi=0.\label{eq:WDquant}
\end{equation}
As before we also expand the log of $\Psi$ and write 
\begin{equation}
\Psi=\exp\left[\frac{i}{\hbar}\left\{ S_{0}+\hbar S_{1}+O\left(\hbar^{2}\right)\right\} \right].\label{eq:Psicorrected}
\end{equation}
Equating the coefficients of $\text{\ensuremath{\hbar^{0}}}$ and
$\hbar$ we get 
\begin{align}
\frac{1}{2}G_{0}^{MN}\frac{\delta S_{0}}{\delta\Phi^{M}}\frac{\delta S_{0}}{\delta\Phi^{N}}+f_{0}[\boldsymbol{\Phi}] & =0\label{eq:HJ0}\\
G_{0}^{MN}\frac{\delta S_{0}}{\delta\Phi^{M}}\frac{\delta S_{1}}{\delta\Phi^{N}} & =\frac{i}{2}G_{0}^{MN}\nabla_{M}\nabla_{N}S_{0}-f_{1}\left[\Phi\right]-\frac{1}{2}G_{1}^{MN}\frac{\delta S_{0}}{\delta\Phi^{M}}\frac{\delta S_{0}}{\delta\Phi^{N}}\label{eq:S1eqnquant}
\end{align}
The first eqn is of course just the original HJ equation, but now
highlighting the fact that the metric and the potential term are just
the classical ones. The second equation has two additional terms on
the LHS compared to eqn. \eqref{eq:prefactoreqn} which come from
the quantum corrections to the Hamiltonian.

Now as before we introduce integral curves defining a trajectory parametrized
by a real valued parameter $s$ in field space (with $C(s)$ reflecting
the arbitrariness of the parametrization), 
\begin{equation}
C(s)\frac{d\Phi^{M}}{ds}=G_{0}^{MN}\frac{\delta S_{0}}{\delta\Phi^{N}}.\label{eq:curve}
\end{equation}
This enables us to integrate \eqref{eq:HJ0} giving 
\begin{equation}
S_{0}\left[\Phi_{s}\right]=-2\int^{s}ds'C\left(s'\right)^{-1}\int_{X}f_{0}\left[\Phi_{s'}\right].\label{S0int}
\end{equation}
with 
\begin{equation}
C(s)=\pm\sqrt{-2\int_{X}f_{0}\left[\Phi_{s}\right]}/\sqrt{\frac{d\Phi^{M}}{ds}G_{MN}\frac{d\Phi^{N}}{ds}}.\label{eq:Cs}
\end{equation}
From \eqref{eq:S1eqnquant} we have 
\begin{equation}
S_{1}\left[\Phi_{s}\right].=\int^{s}ds'C\left(s'\right)^{-1}\int_{X}\left\{ \frac{i}{2}\nabla^{2}S_{0}\left[\Phi_{s'}\right]-f_{1}\left[\Phi_{s'}\right]-\frac{1}{2}\left(G_{1}^{MN}\frac{\delta S_{0}}{\delta\Phi_{s'}^{M}}\frac{\delta S_{0}}{\delta\Phi_{s'}^{N}}\right)\left[\Phi_{s'}\right].\right\} \label{eq:S1}
\end{equation}
To first order quantum correction in $\hbar/i$ times the log of the
wave function is then 
\begin{equation}
S_{0}+\hbar S_{1}=\int^{s}ds'C\left(s'\right)^{-1}\int_{X}\left\{ -2f_{0}\left[\Phi_{s'}\right]+\frac{i\hbar}{2}\nabla^{2}S_{0}\left[\Phi_{s'}\right]-\hbar f_{1}\left[\Phi_{s'}\right]-\frac{\hbar}{2}\left(G_{1}^{MN}\frac{\delta S_{0}}{\delta\Phi_{s'}^{M}}\frac{\delta S_{0}}{\delta\Phi_{s'}^{N}}\right)\left[\Phi_{s'}\right]\right\} .\label{eq:S0S1}
\end{equation}
The second term in the curly brackets above can be shown to be (see
discussion in section \eqref{sec:The-functional-wave} and references
therein),
\begin{align}
\int^{s}ds'C\left(s'\right)^{-1}\int_{X}\frac{i\hbar}{2}\nabla^{2}S_{0}\left[\Phi_{s'}\right] & =-\frac{i\hbar}{2}\ln\det\left[\frac{\delta^{2}S_{0}\left[\Phi_{s}\left(\alpha\right)\right]}{\delta\Phi_{s}^{A}\delta\alpha^{\bar{A}}}\sqrt{G^{-1}}\right]\nonumber \\
 & =\hbar\Gamma_{1}\left[\Phi_{{\rm C}}\right]-\frac{i\hbar}{4}\ln\det G,\label{eq:prefactor-Gamma}
\end{align}
where in the last step we used eqn. \eqref{eq:VVdet}.

Now let us focus on flat space field theory where the metric on field
space is positive definite. For simplicity let us choose it to be
the unit matrix. In this case we have (for a field configuration with
total energy $E$) 
\[
\int_{X}f_{0}\left[\Phi_{s}\right]=\int_{X}\left\{ \left(\nabla_{x}\Phi_{s}\right)^{2}+V\left(\Phi_{s}\right)\right\} -E\equiv U\left[\Phi_{s}\right]-E
\]
The solution for $S$ to order $\hbar$ is, (ignoring the correction
to the field space metric) 
\[
S[\Phi_{s}]=S_{0}+\hbar S_{1}+O\left(\hbar^{2}\right)=\int^{s}ds'C\left(s'\right)^{-1}\int_{X}\left\{ -2\left(f_{0}\left[\Phi_{s'}\right]+\hbar f_{1}\left[\Phi_{s'}\right]\right)\right\} +O\left(\hbar^{2}\right).
\]
Here we've used eqn.\eqref{eq:prefactor-Gamma} and the fact that,
ignoring the kinetic term correction, $\Gamma_{1}=-\int ds'C(s')^{-1}f_{1}$
i.e. the 1 loop correction to the potential in the effective action.
Thus the tunneling wave function (i.e. the WKB factor) is given by
\begin{align}
\Psi & =\exp\left[\frac{i}{\hbar}\left\{ S_{0}+\hbar S_{1}+O\left(\hbar^{2}\right)\right\} \right]\nonumber \\
 & =\exp\left\{ \mp\frac{1}{\hbar}\int^{s}ds'\frac{\sqrt{\frac{d\Phi^{M}}{ds}\delta_{MN}\frac{d\Phi^{N}}{ds}}}{\sqrt{2\int_{X}f_{0}\left[\Phi_{s}\right]}}\int_{X}\left\{ 2\left(f_{0}\left[\Phi_{s'}\right]+\hbar f_{1}\left[\Phi_{s'}\right]\right)\right\} \right\} \label{eq:Psi3}\\
 & =\exp\left\{ \mp\frac{1}{\hbar}\int^{s}ds'\frac{\sqrt{\frac{d\Phi^{M}}{ds}\delta_{MN}\frac{d\Phi^{N}}{ds}}}{\sqrt{2U_{0}\left[\Phi_{s'}\right]}}\left\{ 2\left(U_{0}\left[\Phi_{s'}\right]+\hbar U_{1}\left[\Phi_{s'}\right]\right)\right\} \right\} \label{eq:Psi4}
\end{align}
What this result shows is that the tunneling amplitude can be calculated
as if the potential is the one-loop corrected one (i.e. with $f_{0}\rightarrow f_{0}+\hbar f_{1}$)
except for the fact that the denominator in the integral still only
involves $f_{0}$. It is this that can be computed by finding the
relevant bounce solution as is done in the literature. In other words
(replacing $s'$ by $\tau$) we have, denoting the turning points
by $b,c$ as in figure \eqref{fig:PotentialUFlatSpace} and writing
the total energy as $E=U[\Phi_{A}]$ corresponding to a homogeneous
field configuration sitting at the ``false'' minimum, 
\[
\int_{b}^{c}d\tau\sqrt{2\left(U\left[\Phi_{\tau}\right]-U\left[\Phi_{A}\right]\right)}=S_{E}\left[\Phi_{B}\right]-S_{E}\left[\Phi_{A}\right],\,\,U=U_{0}+\hbar U_{1},
\]
where $S_{E}\left[\Phi_{B}\right]$ $S_{E}\left[\Phi_{A}\right]$
are respectively the Euclidean action of the instanton and background
(i.e. false vacuum) action. Indeed if in the denominator in the exponent
of \eqref{eq:Psi4} we had replaced $U_{0}$ by $U_{0}+\hbar U_{1}$
we would have got exactly the formula that was used to compute the
tunneling rate in the one-loop approximation (see discussion around
eqn. \eqref{eq:lambda_phi_large}. However our systematic discussion
above gives a correction factor to this. Thus the correct formula
for the wave function \eqref{eq:Psi4} can be rewritten as 
\begin{equation}
\Psi=\exp\left\{ \mp\frac{1}{\hbar}\left(S_{E}\left[\Phi_{B}\right]-S_{E}\left[\Phi_{A}\right]\right)\right\} \exp\left\{ \mp\frac{1}{2}\int_{b}^{c}d\tau\frac{2U_{1}}{\sqrt{2U_{0}}}\right\} \label{eq:Psi5}
\end{equation}

This formula is particularly relevant when the one-loop correction
changes the qualitative features of the classical potential as in
the case of the standard model Higgs potential. In this case as is
well known the effective coupling turns negative for large values
of the Higgs field, thus giving a second turning point for the dynamics
of the field. Hence this was calculated by computing the Euclidean
solution for the quantum corrected potential i.e. the instanton corresponding
to tunneling in the quantum corrected potential. This corresponds
to the first factor in \eqref{eq:Psi5}. However we see that there
is a correction term which is of the same order as the one loop correction
term inside the first factor!

In any case it is clear from the above argument that there is no additional
pre-factor to be calculated - in other words the above equation is
correct to $O\left(\hbar\right)$.

Of course this argument is clearly inadequate to justify the actual
estimates done for the lifetime of the SM vacuum since they involve
integrating the beta functions to get the effective potential - rather
than using the one-loop effective action with a couplings at the electro-weak
scale as we've done above. However integrating the beta-functions
is tantamount to summing over leading logs next to leading logs etc.
In effect we have an infinite series in $\hbar$. Clearly in such
a situation it is unclear how to justify the WKB argument which depends
on equating coefficients of powers of $\hbar$, and in practice the
method is usually tractable only to $O\left(\hbar\right)$ (see the
discussion at the beginning of this section).

\subsection{Comments on the literature}

What is done in the literature is to treat $V_{{\rm eff}}=\frac{1}{4}\lambda_{{\rm eff}}\left(\phi\right)\phi^{4}$
with $\lambda_{{\rm eff}}\left(\phi\right)$ obtained by integrating
the beta function equations, as a classical potential - even though
it incorporates an infinite series in $\hbar$. Thus instead of \eqref{eq:WDquant}
we are supposed to use \eqref{eq:S-eqn}, except that now the function
$f$ as well as the metric on field space $G$ are taken to be given
by the RG improved values. In other words the potential and the field
space metric are given by integrating the beta functions from the
electro-weak point to some large scale. Thus (in the flat space case
with just one-scalar field) the Schroedinger equation to be solved
is 
\begin{equation}
\int_{X}\left[-\frac{\hbar^{2}}{2}Z\frac{\delta^{2}}{\delta\phi(x)^{2}}+\frac{1}{2}Z\left(\boldsymbol{\nabla}_{x}\phi\right)^{2}+V_{{\rm eff}}(\phi)\right]\Psi[\phi]=E\Psi[\phi].\label{eq:Seqn_eff}
\end{equation}
where $Z$ and $V_{{\rm eff}}$ are obtained by integrating the relevant
gamma and beta function equations and therefore involve an infinite
series in $\hbar$. However in order to use standard WKB (and hence
Coleman's) arguments to this equation one has to treat $Z$ and $V_{{\rm eff}}$
as if they were classical quantities independent of $\hbar$! If this
is justified then the decay rate is still given by \eqref{eq:lifetime2}
except that the ``classical action'' $S_{E}$ is obtained by solving
the Hamilton-Jacobi equation with this modified Hamiltonian. So the
H-J equation is 
\begin{equation}
\int_{X}\left[\frac{Z}{2}\left(\frac{\partial S}{\partial\phi}\right)^{2}+\frac{Z}{2}\left(\boldsymbol{\nabla}_{x}\phi\right)^{2}+V_{{\rm eff}}(\phi)\right]=E.\label{eq:RG-Hamiltonian}
\end{equation}

Now in calculations of the tunneling rate starting with the work of
Isidori et al \citep{Isidori:2001bm}, it was assumed that once a
possible instability was established by RG running (i.e. the existence
of a scale below $M_{P}$ at which $\lambda_{{\rm eff}}$ goes through
zero to negative values) one could use Coleman's formula under the
assumption that the classical potential is $V=-\frac{|\lambda|}{4}\phi^{4}$
with a constant lambda. In this situation it was argued that there
is still a barrier since even though the potential is negative, since
a bubble is necessarily spatially inhomogeneous so there is gradient
energy and conceivably $U>0$. This turns out to be the case for the
so-called Fubini instanton (see \citep{Andreassen:2016cvx} for a
review and references to the original literature) which minimizes
the Euclidean action, and then the tunneling rate is evaluated at
the point where $|\lambda|$ is maximized (i.e. at the zero of the
beta function). This is then treated as a solution ($S_{0})$ to the
H-J equation and then quantum fluctuations around this minimum (effectively
$S_{1}$) are calculated.

There are however two related issues with this analysis. Firstly if
the potential is negative then clearly the system can roll down classically
as a homogeneous configuration with no barrier. To actually generate
a barrier one really has to assume that a bubble is nucleated. But
this can only happen if there actually is a classical barrier with
$V>0$ in some region as indeed is the case for the standard model.
In that case one needs to confront the situation described above and
treat \eqref{eq:RG-Hamiltonian} as a classical Hamiltonian. Then
one needs to solve the under barrier equations numerically to get
the tunneling exponent $B$.

Now it turns out that for the central values of the Higgs and top
masses the numerical evaluation of $B$ is not very different from
the calculation using the Fubini instanton. But this is a numerical
accident and is extremely sensitive to the initial values of the Higgs
self coupling and the Yukawa coupling of the top. Indeed even the
errors in the experimental values give around ten percent difference
in these calculations. In any case a priori there was no way that
one could know that the constant negative $\lambda$ calculation would
have given a reasonable approximation to the actual numerical calculation,
which of course involved a potential going over many orders of magnitude.
Furthermore the calculations of the prefactor for the Fubini instanton
case \citep{Andreassen:2016cvx}, gave a correction which is of the
same order as the variation in the different estimates of $B$.

In any case even the more accurate numerical calculation uses the
Coleman formula, with the classical action being replaced by one where
the coupling constant is obtained by integrating (numerically) the
beta function, and then evaluating it along a trajectory which solves
the equations of motion. The calculation is of course well-defined.
However the problem is that once the running of the couplings $Z$
and $V_{{\rm eff}}$ are included it is not at all clear how to justify
the WKB procedure - which was the basis for Coleman's derivation of
the tunneling factor as the difference between two classical actions.
The best we can do following standard WKB arguments is what we presented
in the previous subsection.

\subsection{A conjecture}

The discussion in the previous sub-sections suggests the following
speculation for the exact wave functional satisfying the quantum corrected
Schroedinger (or WdW) equation,
\begin{equation}
\Psi\left[\Phi_{\tau}\right]=e^{i\Gamma\left[\Phi_{C}\right]}\Psi[\Phi_{0}]\label{eq:Psiexact}
\end{equation}
with $\Phi_{C}$ a solution of $\delta\Gamma/\Phi=0$ with initial
value $\Phi_{0}$ and final value $\Phi_{\tau}.$ A semi-classical
expansion of $\Gamma$ would be given (for the first two terms) by
equations. \eqref{eq:S0S1},\eqref{eq:prefactor-Gamma}, i.e with
the first order correction being the one-loop correction of the quantum
effective action. The conjecture would be that this holds to all orders.
Needless to say going beyond the leading order corrections (and recursively
correcting the trajectory) is a tall order! But it appears that the
Higgs potential calculation relies on the validity of such an expression
at least to the extent that the couplings in the classical theory
are replaced by the running couplings obtained by integrating the
beta function equations.

\section{Conclusions}

In this note we have rederived the tunneling amplitude for vacuum
decay in flat space in a systematic fashion, starting from the semi-classical
expansion for the time-independent Schroedinger functional differential
equation. We find that while the exponential factor is the same as
that derived by Coleman in his well-known formula, we did not have
to use either the dilute gas approximation or the existence of one
(and only one) negative mode in the determinant of the fluctuation
matrix around the classical solution. Furthermore the details of the
pre-factor are different. Using a wave packet formulation in the corresponding
quantum mechanical situation for the decay of a resonance, we emphasized
the fact that arguing that the effective action has a negative imaginary
part so as to identify the decay width, is both unnecessary and clearly
non-rigorous. Finally we discussed the attempts to use these methods
to calculated the decay of the standard model vacuum and find that
the method used, while seemingly giving plausible answers, lacks a
rigorous derivation.

\section{Acknowledgments}

I wish to thank Sebastian Cespedes, Franceso Muia and Fernando Quevedo
for discussions and collaboration on \citep{Cespedes:2020xpn}, which
work led to the present paper. I also wish to thank Vincenzo Branchina
for discussions on vacuum decay in the standard model.

\section*{Appendix: On analytic continuation of Gaussian integrals}

The original argument of Coleman applied to the situation in the left
hand panel of figure \eqref{fig:PotentialFlatSpace} with a so-called
true vacuum (tv) (i.e. the lower minimum) and a false vacuum (fv)
(i.e. the higher minimum). The argument which depended on Euclidean
QFT was rightly criticized by Andreasson et al \citep{Andreassen:2016cvx},
on the grounds that the starting point i.e. the Euclidean partition
function corresponding to the transition from bounce from fv back
to fv,

\begin{equation}
<A|e^{-HT}|A>=\int_{\phi[-T/2]=A}^{\phi[T/2]=A}[d\phi]e^{-S_{E}[\phi]_{-T/2}^{T/2}},\label{eq:intSE}
\end{equation}
is manifestly real and cannot possibly produce an imaginary part.
The authors then go on to replace Coleman's argument by one which
avoids some of its problems and in particular avoids the use of the
dilute gas approximation. After a rather involved argument they get
an expression for the decay width which is then evaluated in the saddle
point approximation, resulting in the following expression (for tunneling
in QM).

\begin{equation}
\Gamma=\frac{e^{-S_{E}\left[\bar{x}\right]}}{e^{-S_{E}\left[x_{FV}\right]}}\sqrt{S_{E}\left[\bar{x}\right]/m}\left|\frac{1}{\sqrt{\pi}}\Im\left(\frac{\det'S_{E}''\left[\bar{x}\right]}{\det S''\left[x_{FV}\right]}\right)^{-1/2}\right|.\label{eq:AFFS}
\end{equation}
In this relation $\bar{x}$ is the so-called bounce solution corresponding
to a solution which starts at the false vacuum (the higher minimum
in the actual potential but the lower maximum in the inverted potential
which describes motion in Euclidean time) and ends at the same point,
after rolling down and climbing up towards the inverted true vacuum
and then reversing its path and ending up again at the starting point.
$x_{FV}$ on the other hand is the solution where the particle remains
at the false vacuum.

This formula is the same as that derived in \citep{Coleman:1977py}\citep{Callan:1977pt}
and it has the same issue - namely that an imaginary part can only
come from a negative eigenvalue of the operator $S_{E}''\left[\bar{x}\right]$.
Let us trace the origin of this negative eigenvalue. It comes from
evaluating the functional integral for fluctuations around the classical
trajectory $\bar{x}$. The fluctuations are governed by the Schroedinger
operator $-\frac{d^{2}}{dx^{2}}+V''(x)$. This has a zero mode corresponding
to the broken translational invariance of the bounce solution. The
corresponding eigenfunction however has a node, meaning that its eigenvalue
is not the lowest, implying that there is (at least) one negative
eigenvalue. Going back to eqn. \eqref{eq:AFFS} it would seem as if
this will give a negative value to the determinant in the numerator,
hence justifying the replacement of that eqn by

\begin{equation}
\Gamma=\frac{e^{-S_{E}\left[\bar{x}\right]}}{e^{-S_{E}\left[x_{FV}\right]}}\sqrt{S_{E}\left[\bar{x}\right]/m}\left|\frac{1}{\sqrt{\pi}}\left(\frac{\det'S_{E}''\left[\bar{x}\right]}{\det S''\left[x_{FV}\right]}\right)^{-1/2}\right|.\label{eq:AFFS2}
\end{equation}
However the problem is that this is tantamount to declaring that the
inverse Gaussian integral (which would have been the result of having
a negative eigenvalue) instead of being infinite is actually a finite
though imaginary number. This was justified in Coleman's work (for
a detailed discussion see \citep{Coleman:1985rnk,Weinberg:2012pjx})
by arguing that the corresponding integral should extend from negative
infinity to zero and then take off into the imaginary axis. However
this amounts to subtracting an exponentially divergent quantity. The
original integral is over the entire real line and from Cauchy's theorem
one sees that the prescription amounts to subtracting the integral
over the quarter circle in the north-east quadrant which is where
the exponential divergence now resides. Furthermore subtracting an
infinity is inherently ambiguous. 

Let us examine this in detail. Consider the analytic function $e^{\lambda z^{2}},\,\lambda=|\lambda|e^{i\theta},\,0<\theta\leq\pi,z=re^{i\phi}$.
Integrating over a wedge of angle $\phi_{0}$ with one side going
from the origin to $R$ along the real axis, then along a circular
arc from $0$ to $\phi_{0}$ and then back to the origin along the
line $\phi=\phi_{0}$ and using Cauchy's theorem we have 
\begin{align}
\int_{0}^{R}dre^{|\lambda|e^{i\theta}r^{2}} & =-i\int_{0}^{\phi_{0}}d\phi Re^{i\phi}e^{|\lambda|R^{2}\left(\cos\left(2\phi+\theta\right)+i\sin\left(2\phi+\theta\right)\right)}\nonumber \\
 & +e^{i\phi_{0}}\int_{0}^{R}dre^{|\lambda|r^{2}\left(\cos\left(2\phi_{0}+\theta\right)+i\sin\left(2\phi_{0}+\theta\right)\right)}.\label{eq:Cauchy}
\end{align}
The first term will vanish exponentially when $R\rightarrow\infty$,
provided that the cosine is negative, i.e. for 
\begin{equation}
\frac{\pi}{2}<2\phi+\theta\leq\frac{3}{2}\pi.\label{eq:restriction}
\end{equation}
If this is satisfied for $0<\phi<\phi_{0}$, then in the limit we
have 
\[
\int_{0}^{\infty}dre^{|\lambda|e^{i\theta}r^{2}}=e^{i\phi_{0}}\int_{0}^{\infty}dre^{|\lambda|r^{2}\left(\cos\left(2\phi_{0}+\theta\right)+i\sin\left(2\phi_{0}+\theta\right)\right)}.
\]
The RHS of the above is the standard Gaussian integral once we set
$2\phi_{0}+\theta=\pi$ in which case we get 
\begin{equation}
\int_{0}^{\infty}dre^{|\lambda|e^{i\theta}r^{2}}=e^{i\left(\pi-\theta\right)/2}\frac{1}{2}\sqrt{\frac{\pi}{|\lambda|}}\label{eq:generalGaussian}
\end{equation}

\paragraph{Special cases:}

Let us take $\theta=\pi/2.$ As is easily seen all the restrictions
on the angle ranges are satisfied and we have 
\begin{equation}
\int_{0}^{\infty}dre^{i|\lambda|r^{2}}=e^{i\pi/4}\frac{1}{2}\sqrt{\frac{\pi}{|\lambda|}}.\label{eq:imaginaryGaussan}
\end{equation}
Now what the above formula \eqref{eq:AFFS} requires however is the
$\theta=0$ case, i.e. the inverse Gaussian integral! In other words
we have 
\begin{align*}
\int_{0}^{R}dre^{|\lambda|r^{2}} & =-i\int_{0}^{\phi_{0}}d\phi Re^{i\phi}e^{|\lambda|R^{2}\left(\cos\left(2\phi\right)+i\sin\left(2\phi\right)\right)}+\\
 & e^{i\phi_{0}}\int_{0}^{R}dre^{|\lambda|r^{2}\left(\cos\left(2\phi_{0}\right)+i\sin\left(2\phi_{0}\right)\right)}.
\end{align*}
What's done now is to put $\phi_{0}=\pi/2$ to get the second term
on the RHS to be a standard Gaussian. However the first term is clearly
exponentially divergent since in the range $0<\phi<\pi/4$ the cosine
in the exponential is positive. So the above formula is 
\begin{align*}
\int_{0}^{R\rightarrow\infty}dre^{|\lambda|r^{2}} & =\lim_{R\rightarrow\infty}O(e^{|\lambda|R^{2}})-e^{i\pi/2}\sqrt{\frac{\pi}{|\lambda|}}.
\end{align*}
Now the (complex and lambda dependent) divergence is dropped and the
manifestly divergent real integral is replaced by an imaginary quantity!
However subtracting a divergence is an ambiguous procedure so it's
unclear how this can be rigorously justified. The whole procedure
of extracting a imaginary unit from a manifestly real integral (i.e.
the original functional integral \eqref{eq:intSE}) by using the saddle
point approximation appears to be quite mysterious. However it seems
this procedure has been widely accepted in the community for reasons
that are equally mysterious.

In contrast to these ambiguities the derivation in the text is unambiguous.
Of course at the end of the day the dominant factor in the formula
for the decay rate is just the WKB factor and clearly its evaluation
is the same as in the literature (for example it is given by Coleman's
bounce in the appropriate set up). The difference lies in the pre-factor
as we've argued in section \eqref{sec:Flat-space-tunneling}.

  \bibliographystyle{apsrev}
\nocite{*}
\bibliography{myrefs}

\begin{thebibliography}{36}
\expandafter\ifx\csname natexlab\endcsname\relax\def\natexlab#1{#1}\fi
\expandafter\ifx\csname bibnamefont\endcsname\relax
  \def\bibnamefont#1{#1}\fi
\expandafter\ifx\csname bibfnamefont\endcsname\relax
  \def\bibfnamefont#1{#1}\fi
\expandafter\ifx\csname citenamefont\endcsname\relax
  \def\citenamefont#1{#1}\fi
\expandafter\ifx\csname url\endcsname\relax
  \def\url#1{\texttt{#1}}\fi
\expandafter\ifx\csname urlprefix\endcsname\relax\def\urlprefix{URL }\fi
\providecommand{\bibinfo}[2]{#2}
\providecommand{\eprint}[2][]{\url{#2}}

\bibitem[{\citenamefont{Coleman}(1977)}]{Coleman:1977py}
\bibinfo{author}{\bibfnamefont{S.~R.} \bibnamefont{Coleman}},
  \bibinfo{journal}{Phys. Rev.} \textbf{\bibinfo{volume}{D15}},
  \bibinfo{pages}{2929} (\bibinfo{year}{1977}), \bibinfo{note}{[Erratum: Phys.
  Rev.D16,1248(1977)]}.

\bibitem[{\citenamefont{Callan and Coleman}(1977)}]{Callan:1977pt}
\bibinfo{author}{\bibfnamefont{C.~G.} \bibnamefont{Callan}, \bibfnamefont{Jr.}}
  \bibnamefont{and} \bibinfo{author}{\bibfnamefont{S.~R.}
  \bibnamefont{Coleman}}, \bibinfo{journal}{Phys. Rev. D}
  \textbf{\bibinfo{volume}{16}}, \bibinfo{pages}{1762} (\bibinfo{year}{1977}).

\bibitem[{\citenamefont{Coleman and De~Luccia}(1980)}]{Coleman:1980aw}
\bibinfo{author}{\bibfnamefont{S.~R.} \bibnamefont{Coleman}} \bibnamefont{and}
  \bibinfo{author}{\bibfnamefont{F.}~\bibnamefont{De~Luccia}},
  \bibinfo{journal}{Phys. Rev. D} \textbf{\bibinfo{volume}{21}},
  \bibinfo{pages}{3305} (\bibinfo{year}{1980}).

\bibitem[{\citenamefont{Frampton}(1976)}]{Frampton:1976kf}
\bibinfo{author}{\bibfnamefont{P.~H.} \bibnamefont{Frampton}},
  \bibinfo{journal}{Phys. Rev. Lett.} \textbf{\bibinfo{volume}{37}},
  \bibinfo{pages}{1378} (\bibinfo{year}{1976}), \bibinfo{note}{[Erratum:
  Phys.Rev.Lett. 37, 1716 (1976)]}.

\bibitem[{\citenamefont{Frampton}(1977)}]{Frampton:1976pb}
\bibinfo{author}{\bibfnamefont{P.~H.} \bibnamefont{Frampton}},
  \bibinfo{journal}{Phys. Rev. D} \textbf{\bibinfo{volume}{15}},
  \bibinfo{pages}{2922} (\bibinfo{year}{1977}).

\bibitem[{\citenamefont{Cespedes et~al.}(2021)\citenamefont{Cespedes, de~Alwis,
  Muia, and Quevedo}}]{Cespedes:2020xpn}
\bibinfo{author}{\bibfnamefont{S.}~\bibnamefont{Cespedes}},
  \bibinfo{author}{\bibfnamefont{S.~P.} \bibnamefont{de~Alwis}},
  \bibinfo{author}{\bibfnamefont{F.}~\bibnamefont{Muia}}, \bibnamefont{and}
  \bibinfo{author}{\bibfnamefont{F.}~\bibnamefont{Quevedo}},
  \bibinfo{journal}{Phys. Rev. D} \textbf{\bibinfo{volume}{104}},
  \bibinfo{pages}{026013} (\bibinfo{year}{2021}), \eprint{2011.13936}.

\bibitem[{\citenamefont{Andreassen et~al.}(2017)\citenamefont{Andreassen,
  Farhi, Frost, and Schwartz}}]{Andreassen:2016cvx}
\bibinfo{author}{\bibfnamefont{A.}~\bibnamefont{Andreassen}},
  \bibinfo{author}{\bibfnamefont{D.}~\bibnamefont{Farhi}},
  \bibinfo{author}{\bibfnamefont{W.}~\bibnamefont{Frost}}, \bibnamefont{and}
  \bibinfo{author}{\bibfnamefont{M.~D.} \bibnamefont{Schwartz}},
  \bibinfo{journal}{Phys. Rev. D} \textbf{\bibinfo{volume}{95}},
  \bibinfo{pages}{085011} (\bibinfo{year}{2017}), \eprint{1604.06090}.

\bibitem[{\citenamefont{Van~Vleck}(1928)}]{VanVleck:1928zz}
\bibinfo{author}{\bibfnamefont{J.~H.} \bibnamefont{Van~Vleck}},
  \bibinfo{journal}{Proc. Nat. Acad. Sci.} \textbf{\bibinfo{volume}{14}},
  \bibinfo{pages}{178} (\bibinfo{year}{1928}).

\bibitem[{\citenamefont{Brown}(1971)}]{Brown:1971zzc}
\bibinfo{author}{\bibfnamefont{L.~S.} \bibnamefont{Brown}},
  \bibinfo{journal}{American Journal of Physics.}
  \textbf{\bibinfo{volume}{40}}, \bibinfo{pages}{371} (\bibinfo{year}{1971}).

\bibitem[{\citenamefont{Banks and Bender}(1972)}]{Banks:1972xa}
\bibinfo{author}{\bibfnamefont{T.~I.} \bibnamefont{Banks}} \bibnamefont{and}
  \bibinfo{author}{\bibfnamefont{C.~M.} \bibnamefont{Bender}},
  \bibinfo{journal}{J. Math. Phys.} \textbf{\bibinfo{volume}{13}},
  \bibinfo{pages}{1320} (\bibinfo{year}{1972}).

\bibitem[{\citenamefont{Banks et~al.}(1973)\citenamefont{Banks, Bender, and
  Wu}}]{Banks:1973ps}
\bibinfo{author}{\bibfnamefont{T.}~\bibnamefont{Banks}},
  \bibinfo{author}{\bibfnamefont{C.~M.} \bibnamefont{Bender}},
  \bibnamefont{and} \bibinfo{author}{\bibfnamefont{T.~T.} \bibnamefont{Wu}},
  \bibinfo{journal}{Phys. Rev. D} \textbf{\bibinfo{volume}{8}},
  \bibinfo{pages}{3346} (\bibinfo{year}{1973}).

\bibitem[{\citenamefont{Banks and Bender}(1973)}]{Banks:1973uca}
\bibinfo{author}{\bibfnamefont{T.}~\bibnamefont{Banks}} \bibnamefont{and}
  \bibinfo{author}{\bibfnamefont{C.~M.} \bibnamefont{Bender}},
  \bibinfo{journal}{Phys. Rev. D} \textbf{\bibinfo{volume}{8}},
  \bibinfo{pages}{3366} (\bibinfo{year}{1973}).

\bibitem[{\citenamefont{Bitar and Chang}(1978)}]{Bitar:1978vx}
\bibinfo{author}{\bibfnamefont{K.~M.} \bibnamefont{Bitar}} \bibnamefont{and}
  \bibinfo{author}{\bibfnamefont{S.-J.} \bibnamefont{Chang}},
  \bibinfo{journal}{Phys. Rev. D} \textbf{\bibinfo{volume}{18}},
  \bibinfo{pages}{435} (\bibinfo{year}{1978}).

\bibitem[{\citenamefont{Gervais and Sakita}(1977)}]{Gervais:1977nv}
\bibinfo{author}{\bibfnamefont{J.-L.} \bibnamefont{Gervais}} \bibnamefont{and}
  \bibinfo{author}{\bibfnamefont{B.}~\bibnamefont{Sakita}},
  \bibinfo{journal}{Phys. Rev. D} \textbf{\bibinfo{volume}{16}},
  \bibinfo{pages}{3507} (\bibinfo{year}{1977}).

\bibitem[{\citenamefont{Tanaka et~al.}(1994)\citenamefont{Tanaka, Sasaki, and
  Yamamoto}}]{Tanaka:1993ez}
\bibinfo{author}{\bibfnamefont{T.}~\bibnamefont{Tanaka}},
  \bibinfo{author}{\bibfnamefont{M.}~\bibnamefont{Sasaki}}, \bibnamefont{and}
  \bibinfo{author}{\bibfnamefont{K.}~\bibnamefont{Yamamoto}},
  \bibinfo{journal}{Phys. Rev. D} \textbf{\bibinfo{volume}{49}},
  \bibinfo{pages}{1039} (\bibinfo{year}{1994}).

\bibitem[{\citenamefont{Poisson}(2009)}]{Poisson:2009pwt}
\bibinfo{author}{\bibfnamefont{E.}~\bibnamefont{Poisson}},
  \emph{\bibinfo{title}{{A Relativist's Toolkit: The Mathematics of Black-Hole
  Mechanics}}} (\bibinfo{publisher}{Cambridge University Press},
  \bibinfo{year}{2009}).

\bibitem[{\citenamefont{DeWitt-Morette}(1976)}]{DeWitt-Morette:1976ydh}
\bibinfo{author}{\bibfnamefont{C.}~\bibnamefont{DeWitt-Morette}},
  \bibinfo{journal}{Annals Phys.} \textbf{\bibinfo{volume}{97}},
  \bibinfo{pages}{367} (\bibinfo{year}{1976}), \bibinfo{note}{[Erratum: Annals
  Phys. 101, 682 (1976)]}.

\bibitem[{\citenamefont{DeWitt}(2003)}]{DeWitt:2003pm}
\bibinfo{author}{\bibfnamefont{B.~S.} \bibnamefont{DeWitt}},
  \emph{\bibinfo{title}{{The global approach to quantum field theory. Vol. 1,
  2}}}, vol. \bibinfo{volume}{114} (\bibinfo{publisher}{Int. Ser. Monogr.
  Phys.}, \bibinfo{year}{2003}).

\bibitem[{\citenamefont{Merzbacher}(1998)}]{Merzbacher:1998}
\bibinfo{author}{\bibfnamefont{E.}~\bibnamefont{Merzbacher}},
  \emph{\bibinfo{title}{{Quantum Mechanics third edition}}}
  (\bibinfo{publisher}{Wiley}, \bibinfo{year}{1998}), ISBN
  \bibinfo{isbn}{0-471-88702-1}.

\bibitem[{\citenamefont{Coleman et~al.}(1978)\citenamefont{Coleman, Glaser, and
  Martin}}]{Coleman:1977th}
\bibinfo{author}{\bibfnamefont{S.~R.} \bibnamefont{Coleman}},
  \bibinfo{author}{\bibfnamefont{V.}~\bibnamefont{Glaser}}, \bibnamefont{and}
  \bibinfo{author}{\bibfnamefont{A.}~\bibnamefont{Martin}},
  \bibinfo{journal}{Commun. Math. Phys.} \textbf{\bibinfo{volume}{58}},
  \bibinfo{pages}{211} (\bibinfo{year}{1978}).

\bibitem[{\citenamefont{Markkanen et~al.}(2018)\citenamefont{Markkanen,
  Rajantie, and Stopyra}}]{Markkanen:2018pdo}
\bibinfo{author}{\bibfnamefont{T.}~\bibnamefont{Markkanen}},
  \bibinfo{author}{\bibfnamefont{A.}~\bibnamefont{Rajantie}}, \bibnamefont{and}
  \bibinfo{author}{\bibfnamefont{S.}~\bibnamefont{Stopyra}},
  \bibinfo{journal}{Front. Astron. Space Sci.} \textbf{\bibinfo{volume}{5}},
  \bibinfo{pages}{40} (\bibinfo{year}{2018}), \eprint{1809.06923}.

\bibitem[{\citenamefont{Krive and Linde}(1976)}]{Krive:1976sg}
\bibinfo{author}{\bibfnamefont{I.~V.} \bibnamefont{Krive}} \bibnamefont{and}
  \bibinfo{author}{\bibfnamefont{A.~D.} \bibnamefont{Linde}},
  \bibinfo{journal}{Nucl. Phys. B} \textbf{\bibinfo{volume}{117}},
  \bibinfo{pages}{265} (\bibinfo{year}{1976}).

\bibitem[{\citenamefont{Degrassi et~al.}(2012)\citenamefont{Degrassi, Di~Vita,
  Elias-Miro, Espinosa, Giudice, Isidori, and Strumia}}]{Degrassi:2012ry}
\bibinfo{author}{\bibfnamefont{G.}~\bibnamefont{Degrassi}},
  \bibinfo{author}{\bibfnamefont{S.}~\bibnamefont{Di~Vita}},
  \bibinfo{author}{\bibfnamefont{J.}~\bibnamefont{Elias-Miro}},
  \bibinfo{author}{\bibfnamefont{J.~R.} \bibnamefont{Espinosa}},
  \bibinfo{author}{\bibfnamefont{G.~F.} \bibnamefont{Giudice}},
  \bibinfo{author}{\bibfnamefont{G.}~\bibnamefont{Isidori}}, \bibnamefont{and}
  \bibinfo{author}{\bibfnamefont{A.}~\bibnamefont{Strumia}},
  \bibinfo{journal}{JHEP} \textbf{\bibinfo{volume}{08}}, \bibinfo{pages}{098}
  (\bibinfo{year}{2012}), \eprint{1205.6497}.

\bibitem[{\citenamefont{Branchina et~al.}(2015)\citenamefont{Branchina,
  Messina, and Sher}}]{Branchina:2014rva}
\bibinfo{author}{\bibfnamefont{V.}~\bibnamefont{Branchina}},
  \bibinfo{author}{\bibfnamefont{E.}~\bibnamefont{Messina}}, \bibnamefont{and}
  \bibinfo{author}{\bibfnamefont{M.}~\bibnamefont{Sher}},
  \bibinfo{journal}{Phys. Rev. D} \textbf{\bibinfo{volume}{91}},
  \bibinfo{pages}{013003} (\bibinfo{year}{2015}), \eprint{1408.5302}.

\bibitem[{\citenamefont{Isidori et~al.}(2001)\citenamefont{Isidori, Ridolfi,
  and Strumia}}]{Isidori:2001bm}
\bibinfo{author}{\bibfnamefont{G.}~\bibnamefont{Isidori}},
  \bibinfo{author}{\bibfnamefont{G.}~\bibnamefont{Ridolfi}}, \bibnamefont{and}
  \bibinfo{author}{\bibfnamefont{A.}~\bibnamefont{Strumia}},
  \bibinfo{journal}{Nucl. Phys. B} \textbf{\bibinfo{volume}{609}},
  \bibinfo{pages}{387} (\bibinfo{year}{2001}), \eprint{hep-ph/0104016}.

\bibitem[{\citenamefont{Coleman}(1985)}]{Coleman:1985rnk}
\bibinfo{author}{\bibfnamefont{S.}~\bibnamefont{Coleman}},
  \emph{\bibinfo{title}{{Aspects of Symmetry}: {Selected Erice Lectures}}}
  (\bibinfo{publisher}{Cambridge University Press},
  \bibinfo{address}{Cambridge, U.K.}, \bibinfo{year}{1985}), ISBN
  \bibinfo{isbn}{978-0-521-31827-3}.

\bibitem[{\citenamefont{Weinberg}(2012)}]{Weinberg:2012pjx}
\bibinfo{author}{\bibfnamefont{E.~J.} \bibnamefont{Weinberg}},
  \emph{\bibinfo{title}{{Classical solutions in quantum field theory}:
  {Solitons and Instantons in High Energy Physics}}}, Cambridge Monographs on
  Mathematical Physics (\bibinfo{publisher}{Cambridge University Press},
  \bibinfo{year}{2012}), ISBN \bibinfo{isbn}{978-0-521-11463-9,
  978-1-139-57461-7, 978-0-521-11463-9, 978-1-107-43805-7}.

\bibitem[{\citenamefont{Espinosa}(2019)}]{Espinosa:2019hbm}
\bibinfo{author}{\bibfnamefont{J.~R.} \bibnamefont{Espinosa}},
  \bibinfo{journal}{Phys. Rev. D} \textbf{\bibinfo{volume}{100}},
  \bibinfo{pages}{105002} (\bibinfo{year}{2019}), \eprint{1908.01730}.

\bibitem[{\citenamefont{Espinosa}(2020)}]{Espinosa:2020qtq}
\bibinfo{author}{\bibfnamefont{J.~R.} \bibnamefont{Espinosa}},
  \bibinfo{journal}{JCAP} \textbf{\bibinfo{volume}{06}}, \bibinfo{pages}{052}
  (\bibinfo{year}{2020}), \eprint{2003.06219}.

\bibitem[{\citenamefont{Bando et~al.}(1993)\citenamefont{Bando, Kugo, Maekawa,
  and Nakano}}]{Bando:1992wy}
\bibinfo{author}{\bibfnamefont{M.}~\bibnamefont{Bando}},
  \bibinfo{author}{\bibfnamefont{T.}~\bibnamefont{Kugo}},
  \bibinfo{author}{\bibfnamefont{N.}~\bibnamefont{Maekawa}}, \bibnamefont{and}
  \bibinfo{author}{\bibfnamefont{H.}~\bibnamefont{Nakano}},
  \bibinfo{journal}{Prog. Theor. Phys.} \textbf{\bibinfo{volume}{90}},
  \bibinfo{pages}{405} (\bibinfo{year}{1993}), \eprint{hep-ph/9210229}.

\bibitem[{\citenamefont{Ai et~al.}(2019)\citenamefont{Ai, Garbrecht, and
  Tamarit}}]{Ai:2019fri}
\bibinfo{author}{\bibfnamefont{W.-Y.} \bibnamefont{Ai}},
  \bibinfo{author}{\bibfnamefont{B.}~\bibnamefont{Garbrecht}},
  \bibnamefont{and} \bibinfo{author}{\bibfnamefont{C.}~\bibnamefont{Tamarit}},
  \bibinfo{journal}{JHEP} \textbf{\bibinfo{volume}{12}}, \bibinfo{pages}{095}
  (\bibinfo{year}{2019}), \eprint{1905.04236}.

\bibitem[{\citenamefont{Bachlechner}(2016)}]{Bachlechner:2016mtp}
\bibinfo{author}{\bibfnamefont{T.~C.} \bibnamefont{Bachlechner}},
  \bibinfo{journal}{JHEP} \textbf{\bibinfo{volume}{12}}, \bibinfo{pages}{155}
  (\bibinfo{year}{2016}), \eprint{1608.07576}.

\bibitem[{\citenamefont{Susskind}(2021)}]{Susskind:2021yvs}
\bibinfo{author}{\bibfnamefont{L.}~\bibnamefont{Susskind}}
  (\bibinfo{year}{2021}), \eprint{2107.11688}.

\bibitem[{\citenamefont{De~Alwis et~al.}(2020)\citenamefont{De~Alwis, Muia,
  Pasquarella, and Quevedo}}]{DeAlwis:2019rxg}
\bibinfo{author}{\bibfnamefont{S.~P.} \bibnamefont{De~Alwis}},
  \bibinfo{author}{\bibfnamefont{F.}~\bibnamefont{Muia}},
  \bibinfo{author}{\bibfnamefont{V.}~\bibnamefont{Pasquarella}},
  \bibnamefont{and} \bibinfo{author}{\bibfnamefont{F.}~\bibnamefont{Quevedo}},
  \bibinfo{journal}{Fortsch. Phys.} \textbf{\bibinfo{volume}{68}},
  \bibinfo{pages}{2000069} (\bibinfo{year}{2020}), \eprint{1909.01975}.

\bibitem[{\citenamefont{Farhi et~al.}(1990)\citenamefont{Farhi, Guth, and
  Guven}}]{Farhi:1989yr}
\bibinfo{author}{\bibfnamefont{E.}~\bibnamefont{Farhi}},
  \bibinfo{author}{\bibfnamefont{A.~H.} \bibnamefont{Guth}}, \bibnamefont{and}
  \bibinfo{author}{\bibfnamefont{J.}~\bibnamefont{Guven}},
  \bibinfo{journal}{Nucl. Phys. B} \textbf{\bibinfo{volume}{339}},
  \bibinfo{pages}{417} (\bibinfo{year}{1990}).

\bibitem[{\citenamefont{Fischler et~al.}(1990)\citenamefont{Fischler, Morgan,
  and Polchinski}}]{Fischler:1990pk}
\bibinfo{author}{\bibfnamefont{W.}~\bibnamefont{Fischler}},
  \bibinfo{author}{\bibfnamefont{D.}~\bibnamefont{Morgan}}, \bibnamefont{and}
  \bibinfo{author}{\bibfnamefont{J.}~\bibnamefont{Polchinski}},
  \bibinfo{journal}{Phys. Rev. D} \textbf{\bibinfo{volume}{42}},
  \bibinfo{pages}{4042} (\bibinfo{year}{1990}).

\end{thebibliography}

\end{document}